\documentstyle[11pt]{article}
\input shorts.sty
\margins

\input epsf

\title{A necessary and sufficient instability condition for\\
inviscid shear flow}

\author{N.~J.~Balmforth
%\thanks{Address for correspondence:
%Department of Theoretical Mechanics,
%University of Nottingham, NG7 2RD, UK}
\\
{\small and}\\
P.~J.~Morrison
\thanks{Address for correspondence:
Department of Physics, C1600,
University of Texas, Austin, TX 78712}\\
%\thanks{Electronic mail: morrison@hagar.ph.utexas.edu}\\
%\physifspop
}

\date{February 1998}

\def\eps{{\epsilon}}
\def\calp{{\cal P}}

\def\ltwid{\mathrel{\raise.3ex\hbox{$<$\kern-.75em\lower1ex\hbox{$\sim$}}}}
\def\gtwid{\mathrel{\raise.3ex\hbox{$>$\kern-.75em\lower1ex\hbox{$\sim$}}}}

\def\part{{\partial}}
\def\ome{{\omega}}
\def\cag{{\cal G}}
\def\cah{{\cal H}}
\def\eps{{\epsilon}}
\def\cad{{\cal D}}

\def\cac{{\cal C}}

\def\citenum#1{{\def\@cite##1##2{##1}\cite{#1}}}

\bibliographystyle{pfa}

\begindoc

\maketitle

\hrule
\bigskip

\begin{abstract}
We derive a condition that is  necessary and sufficient
for the instability of inviscid, two-dimensional, plane parallel,
shear flow with equilibrium velocity profiles that are
monotonic, real analytic, functions of the cross stream
coordinate.  The analysis, which is based upon the Nyquist
method,  includes  a means for delineating  the possible
kinds of bifurcations  that involve the presence of the
continuous spectrum,  including those that occur at nonzero
wavenumber.  Several examples are given.

\bigskip

\end{abstract}

\hrule

\section{Introduction}

The linear stability of inviscid, incompressible, two-dimensional,
plane parallel, shear flow was considered over  a century ago by
Rayleigh, Kelvin, and others. A principal result on the subject is
Rayleigh's celebrated inflection point theorem \citenum{R80},
which states that for an equilibrium flow to be unstable, the
equilibrium velocity profile must contain an inflection point.
That is, if the velocity  {\sl profile} is given by $U(y)$, where
$y$ is the cross-stream coordinate, then there must be a point, $y=y_I$,
for which $U''(y_I)=0$. Much later, in 1950, Fj{\o}rtoft \citenum{F50}
generalized the theorem by showing that, moreover, if there is one inflection
point, then $U'''(y_I)/U'(y_I)<0$ is required for instability (see
\citenum{Bar} for further extensions). Both Rayleigh's Theorem and
Fj{\o}rtoft's subsequent generalization are necessary conditions for
instability, but they are {\sl not} sufficient. That is, even
though an equilibrium profile  may contain a vorticity minimum, it is
not necessarily unstable. The point of this paper is to derive,
for a large class of equilibrium  velocity  profiles, a condition
that is   {\sl necessary and sufficient} for instability.

\subsection{Overview}
\label{overview}

The procedure we use to derive the instability condition is
inspired by techniques developed for the Vlasov equation. For
that problem one can find a condition that is  necessary
and sufficient for instability by using the Nyquist method, a
method that leads to what is known  as the Penrose criterion
\citenum{P60} in plasma physics.  The Penrose criterion follows
fairly straightforwardly for the linear  Vlasov problem because
the discrete eigenvalues satisfy an explicit   dispersion
relation. However, in  the context of the Euler   equation
governing the shearing fluid, the  relevant  eigenvalue
problem leads to Rayleigh's equation. This equation cannot
be manipulated into an explicit dispersion relation, and a
sufficient condition for instability of velocity profiles of a
general form  has not previously been given. We note,
though, that \citenum{RS64} and \citenum{TU81} present a
sufficient condition for long wave instability, and in
\citenum{RS64} 
a necessary and sufficient condition  for
instability of profiles with a single inflection point (where
bifurcations are restricted to occur through zero wavenumber)  was
obtained.

Here we derive a  condition that is  necessary and
sufficient for instability for a class of  velocity
profiles, $U(y)$, where $y$ is the cross stream coordinate.
Specifically, we consider  profiles that are
monotonic functions when $y\in[-1,1]$, which we refer to as the
``flow domain,"  and which are real analytic. This latter limitation
means that $U(y)$ has a convergent Taylor series on $[-1,1]$ and
thus possesses an analytic continuation into the complex $y$-plane.
Hence, there exists a  neighborhood (an open set of the
complex plane) that contains the interval $[-1,1]$ in which $U(y)$
is analytic and in which $U'(y)\neq 0$. We assume this neighborhood
is as large as is needed in subsequent calculations.  
%The
%limitation of analyticity is not especially restrictive, since
%according to the Weierstrass approximation theorem we can
%approximate any continuous function arbitrarily well with a
%polynomial. 
Both the monotonicity and  analyticity limitations can
be generalized, but  we will not attempt this
here.

\subsection{Summary of result}
\label{summary}

In the remainder of this section we summarize our main result,
which amounts to a prescription for obtaining the condition that is
necessary and sufficient for instability: given the solution,
$\psi(y,c_r)$, to the Fredholm integral equation,
\bq
\psi(y,c_r) =
\cag(y,y_c) +
\int_{-1}^1{ \cag(y,y')-\cag(y,y_c) \over U(y')-c_r}U''(y')
\psi(y',c_r)dy'
,
\label{Fred}
\eq
where $c_r=U(y_c)$ and
$\cag(y,y')$ is a Green function (given in Eq.~\ref{2.5} below)
containing the streamwise wavenumber
$k$ and embodying the boundary conditions,
we construct the ``Nyquist function,''
\bq
\epsilon(c_r) =
1 - \calp \int_{-1}^1 {U'' (y) \psi(y,c_r)\over U(y)-c_r}dy
- i\pi{U''(y_c)\psi(y_c,c_r)\over U'(y_c)}
,
\label{one}
\eq
where $\calp$ denotes the Cauchy principal value.
We then plot $\epsilon(c_r)$ on the
$(\epsilon_r,\epsilon_i)-$plane for $c_r$ running from
$U(-1)$ to $U(1)$, or, equivalently, for $y_c$ along the interval
$[-1,1]$. The profile $U(y)$ is exponentially unstable for that value of
$k$ used in Eq.~(\ref{Fred}) if and only if the resulting
path loops around the origin, $\epsilon_r=\epsilon_i=0$.
In fact, the path can only cross the $\epsilon_r-$axis at
the {\sl inflection points}, $y_I$, of the velocity profile
where $U''(y_I)=0$.
This leads to the
following  necessary and sufficient condition for
instability:  $\epsilon_r(c_I)<0$ with $c_I=U(y_I)$,
or equivalently
\bq
\int_{-1}^1 {U''(y)\psi(y,c_I)\over U(y)-c_I} dy > 1
,
\label{two}
\eq
for one of the $y_I$'s.

In Section~\ref{num}, among other examples, we treat the profile
$U(y) = \tanh (\beta y)$ and arrive at the ``Nyquist plots'' shown
in Fig.~\ref{fig:nf1}. For this equilibrium profile with $\beta=2$
the plots of Fig.~\ref{fig:nf1}(a) loop around the origin  and
$\epsilon_r(c_I)<0$ (where $c_I=0$) provided $k\ltwid 1.832$. Hence for $\beta=2$ the
profile is
unstable over the band of wavenumbers $0<k\ltwid1.832$.
Figure \ref{fig:nf1}(b) illustrates the onset of instability,
which occurs through $k=0$,  as the parameter $\beta$ is changed.

\begin{figure}
\begin{center}
\leavevmode
\epsfysize=7.cm
\epsfbox{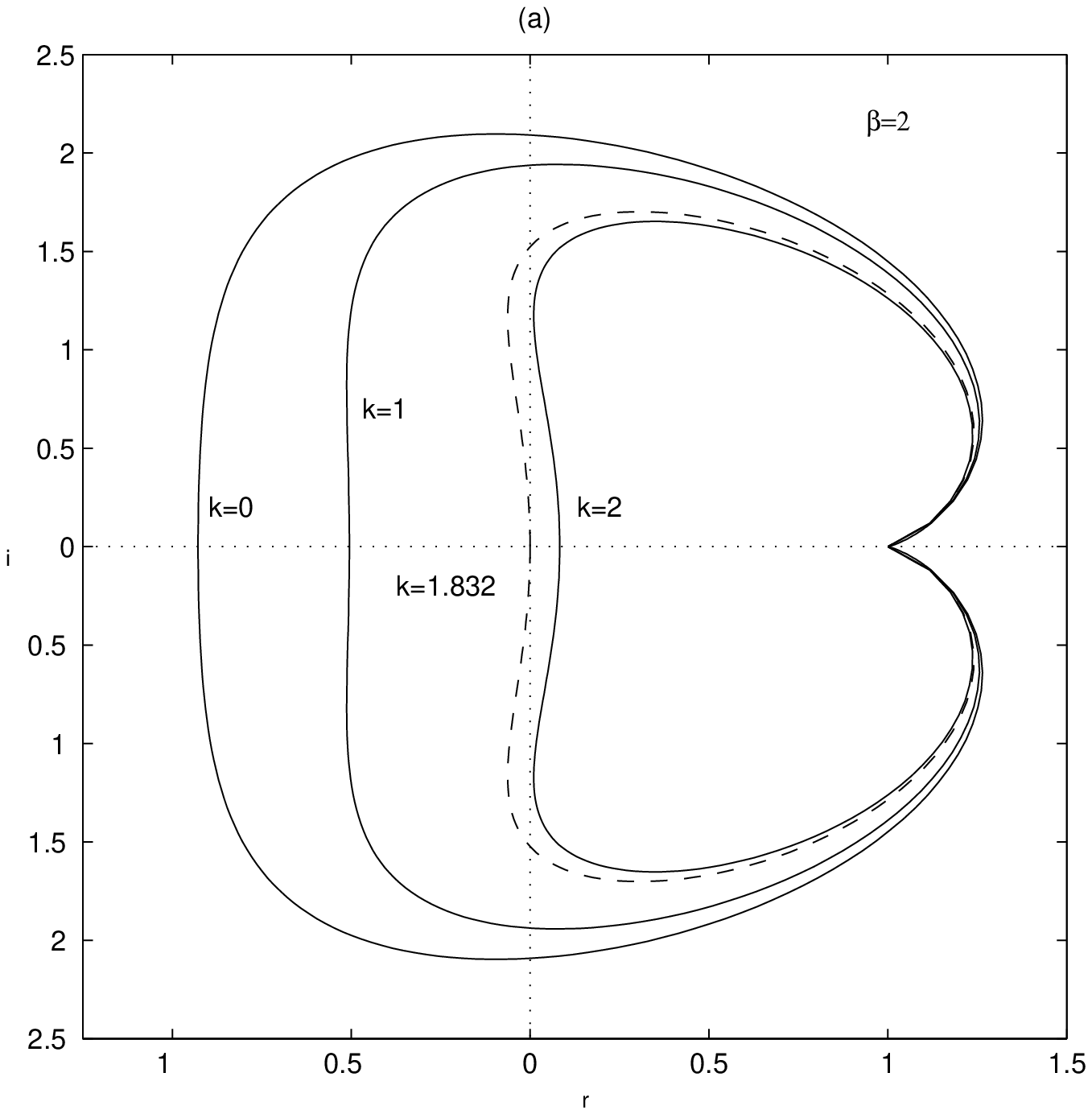}
\epsfysize=7.cm
\epsfbox{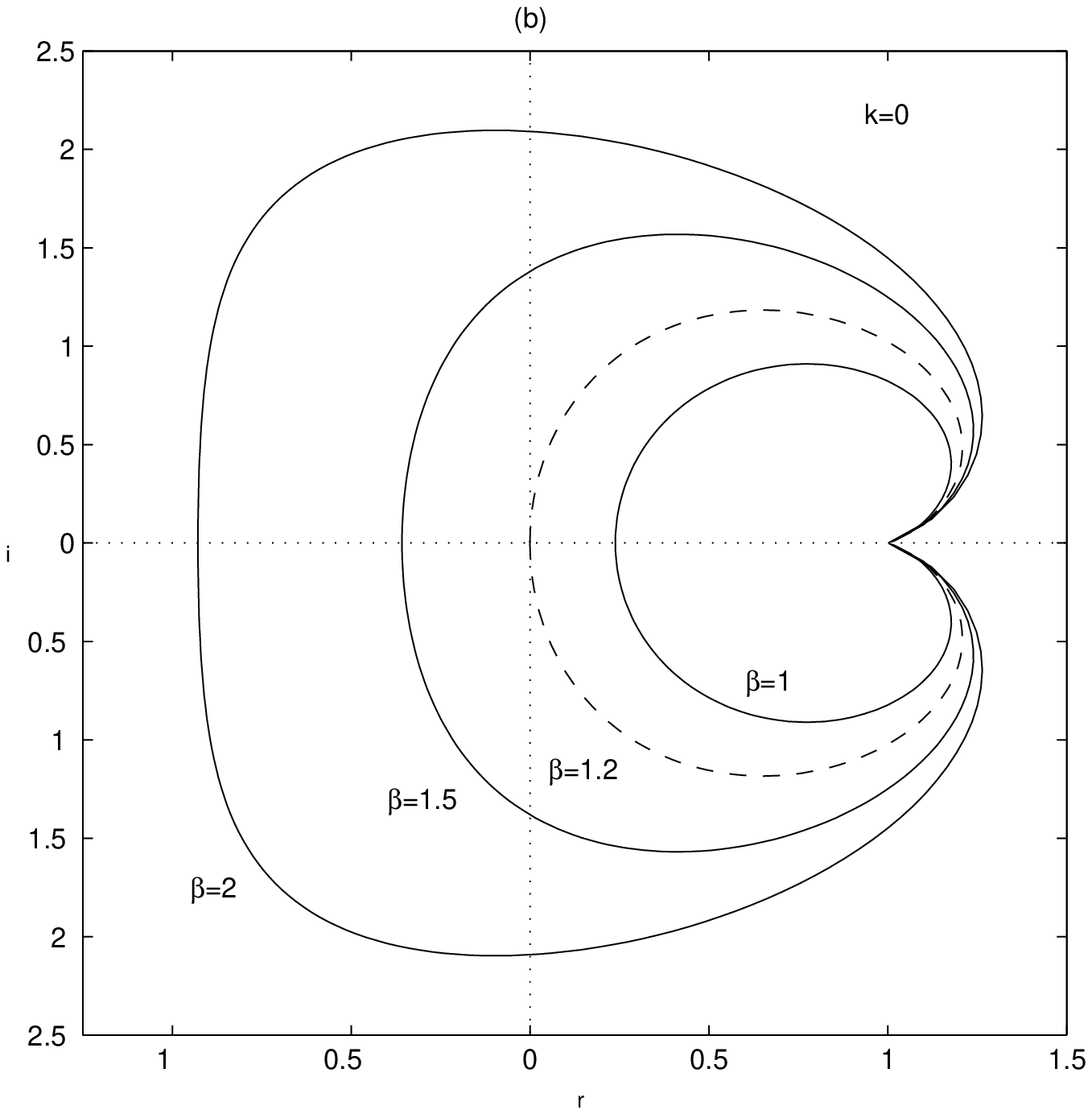}
\end{center}
\caption{Nyquist plots for the single inflection point profile,
$U(y)=\tanh\beta y$. (a) Four plots for $\beta=2$ and $k=0$, $1$, and $2$,
and  the critical value for the onset of instability,
$k=k_c \simeq 1.832$ (dashed curve). (b) Four plots for $k=0$ and $\beta=1$,
$1.5$, and $2$,  and  the critical value for the onset of instability,
$\beta=\beta_c \simeq 1.2$ (dashed curve).}
\label{fig:nf1}
\end{figure}

In order to prove the result outlined above we need various
mathematical results concerning Rayleigh's equation. These are derived in
Sections~\ref{rev}--\ref{nyq}. Application to specific equilibrium velocity
profiles is presented in Sections~\ref{special} and \ref{num}. Finally, in
Section~\ref{con} we   summarize,  place our  work in context, and discuss future
developments.

\section{Review}
\label{rev}

\subsection{Formulation}

For a two-dimensional, inviscid fluid contained within the channel,
$x\in(-\infty,\infty)$ and $y\in[-1,1]$, an equilibrium state is given by
any flow, $(U,0)$, where $U=U(y)$ is the equilibrium velocity  profile.
Infinitesimal disturbances to such an equilibrium are described by the
equation \citenum{R80},
\bq
\part_t \tilde\ome + U(y) \part_x \tilde \ome = U''(y) \part_x\tilde\psi ,
\label{2.1}
\eq
where $\tilde\ome(x,y,t)$ is the vorticity perturbation,
$\tilde\psi(x,y,t)$ is the associated streamfunction, which are related by
\bq
\tilde\ome = \nabla^2 \tilde\psi ,
\label{2.2}
\eq
and the boundary conditions are $\tilde\psi(x,\pm1,t)=0$.  Throughout
this paper we will consider solutions of the form
$\tilde\ome=\ome(k,y,t)\exp(ikx)$ and
$\tilde\psi=\psi(k,y,t)\exp(ikx)$, in which case (\ref{2.1}) becomes
\bq
\part_t \ome + ik U(y)  \ome = i k U''(y) \psi\, ,
\label{2.2a}
\eq
and the inverse of (\ref{2.2}) takes the form,
\bq
\psi(y,t) = \int_{-1}^1 \cag(y,y') \ome(y',t)dy' , \label{2.4}
\eq
where
\bq
\cag(y,y')= \left\{ \matrix{
-\sinh k(1-y) \sinh k(1+y') / k \sinh 2k &\,
{\rm if\ } \, y > y' \cr
-\sinh k(1+y) \sinh k(1-y') / k \sinh 2k & \, {\rm if\ }
 \,  y < y' \,. }
\right.
\label{2.5}
\eq
Here and henceforth we suppress the $k$-dependence in the arguments of all
functions.

\subsection{Laplace transform}

We need some results associated with the Laplace
transform approach to the solution of (\ref{2.2a}) as an initial value
problem; so we give a brief review of this approach here.  More detailed
accounts are given in \citenum{C60}, \citenum{DH66}, and
with considerably  more rigor  in \citenum{RS66}.

Recall, the Laplace transform pair is
\bq
\Psi(y,p)= \int_0^{\infty} e^{-pt} \, \psi(y,t)\, dt
\,,\quad\quad
\psi(y,t)= \frac{1}{2\pi i} \int_C e^{pt}\,  \Psi(y,p)\, dp \,,
\label{2.5a}
\eq
where $C$ is the Bromwich contour that runs parallel to the imaginary
$p$-axis  and to the right of all singularities of the  integrand. Using
the above, (\ref{2.2a}) becomes
\bq
\left(\frac{\partial^2}{\partial y^2} - k^2
- \frac{ik U''}{p + ikU}\right) \Psi(y,p)
= \frac{\omega(y,0)}{p + ikU}
,
\label{2.5b}
\eq
where $\omega(y,0)$, the initial value of the vorticity, satisfies
$\omega(y,0) = (\partial^2 /\partial y^2 -k^2 )\psi(y,0)$.  The
solution  to (\ref{2.5}) can be written formally as
\bq
\Psi(y,p)=\int_{-1}^{1} \cah(y,y';p) \,
\frac{\omega(y',0)}{p + ikU(y')}\, dy'\,,
\label{2.5c}
\eq
where the Green function, $\cah(y,y';p)$, is given by
\bq
\cah(y,y';p)= \left\{
\matrix{
- \Psi_1(y,p)\,\Psi_2(y',p)/ W(\Psi_1,\Psi_2)  &\,\,\,
{\rm if\ } \, -1 \leq  y \leq y' \cr
- \Psi_1(y',p)\,\Psi_2(y,p)/ W(\Psi_1,\Psi_2)  &\,
{\rm if\ }\, \,  \, y' \leq  y \leq 1 \,,}
\right.
\label{2.5d}
\eq
with the Wronskian
\bq
W(\Psi_1,\Psi_2):= \Psi_1(y,p)\,\Psi'_2(y,p)
-\Psi'_1(y,p)\,\Psi_2(y,p)\,.
\label{2.5e}
\eq
Here $\Psi_1$ and $\Psi_2$ are any two solutions to (\ref{2.5b}) with
the right-hand side set to zero and that satisfy the boundary conditions,
$\Psi_1(-1,p)= \Psi_2(1,p)=0$. Since (\ref{2.5b})  does not contain a
term involving the first derivative with respect to $y$, it is an
elementary result (see \citenum{I44}, Chap.~V) that the Wronskian is
independent of $y$. Arguments pertaining to  the inverse
Laplace transform of (\ref{2.5c}) imply that the dispersion relation for
discrete normal modes is given by
\bq
W(\Psi_1,\Psi_2)=0\,,
\label{2.5f}
\eq
with ${\rm Re}(p)>0$.

\subsection{Normal modes}

An alternative approach to Laplace transformation follows when
we search for normal modes at the outset by assuming
that $\ome(y,t)=\ome(y)\exp(-ikct)$ and
$\psi(y,t)=\psi(y)\exp(-ikct)$. Inserting these expressions into
(\ref{2.2a}) yields  Rayleigh's equation,
\bq
(U-c) \ome = (U-c) (\psi''-k^2\psi) = U''\psi\, .
\label{2.3}
\eq
which is an  equation for the eigenfunction $\psi(y)$
corresponding to the eigenvalue $c$. In the next section we describe some
properties of this equation, for both real and imaginary values of $c$,
that we will need later.

\section{Rayleigh's equation}
\label{form}

\subsection{Singular points and solutions in the complex plane}

We will need several facts about the solutions to Rayleigh's
equation (\ref{2.3}). Firstly, the point $y_c$, for which
$U(y_c)=c$, is a singular point of (\ref{2.3}). If $c$ is real
then this point is commonly referred to as a  ``critical
layer.''
%and we will use this terminology for this singular
%point even when $c$ is complex and  the singular point  occurs
%in the complex $y$-plane.
Since $U(y)$ is a monotonic function for
$y\in[-1,1]$ there is at most one such critical layer in the
flow domain, and this occurs when $c$ is in the range of
$U(y)$. Because of monotonicity and real analyticity of $U$ for
$y$ in the flow domain, there exists  a neighborhood of
$[-1,1]$ in the complex plane in which  the only  singular
point of (\ref{2.3}) is the critical layer.  We add that,
about any point of the flow domain,  there  exists a
neighborhood in which the complex variables
$y_c$ and $c$ are in one-to-one correspondence. Without loss of
generality, we will take $U'(y)>0$ when $y\in[-1,1]$ for
definiteness.

Rayleigh's equation has a singular point at the critical layer,
but this is not the only possible singular point. First,
there may be other points in the complex $y$ plane for which
$U(y)=c$. Second,
although no others exist when $U$ is continued into  a
neighborhood of $[-1,1]$, it is likely that additional singular
points occur in $U$ when that function
is further analytically continued into the complex plane.
These singularities can then show up in Rayleigh's equation
(though it is worth noting that because $U$ occurs in Rayleigh's equation
only in  the coefficient $U''/(U-c)$, for  meromorphic $U$ all
singular points in the finite complex plane are regular;
this can easily be shown by Laurent expansion). However, here
we will only need to consider the critical layer.

{}From the
elementary theory of ordinary differential equations in the complex
plane (see {\it e.g.}\ \citenum{WW27}, Chap.~X) it is known that there
are two analytic solutions for $y$ in a neighborhood of any ordinary
point. Moreover, in general, branch points occur in the solution at
the locations of regular singular points. If we fix
the two parameters, $k$ and $c$,
then the locations of the singular points are fixed for a given
equilibrium profile.

In a neighborhood of the critical point in the complex $y$-plane
it is straightforward to obtain two Frobenius series solutions of the
following form:
\bq
\psi_g(y,c) = [U(y)-c] \varphi_1(y,c)
\label{2.6}
\eq
and
\bq
\psi_b(y,c) = [U(y)-c] \log |U(y)-c| \varphi_2(y,c) +
\varphi_3(y,c) ,
\label{2.7}
\eq
where $\varphi_1$, $\varphi_2$, and $\varphi_3$ are analytic
for $y$ in this neighborhood  and for $c$ in the finite complex
$c$-plane.
%(Note,  we  use the  loose  notation  where we interchange
%dependence upon $c$ and $y_c$; for example,
%$\psi(y,y_c)=\psi(y,U(y_c))=\psi(y,c)$.)

%A straightforward way  to see the analyticity in
%$c$ (except for the  critical layer singularity) and to obtain
%the equations above, is to transform Rayleigh's equation to the
%variable $u:=U(y)$. The analytic extension of this
%transformation maps a region of the complex $y$-plane to a
%region of the complex $u$-plane.  For $y$ in a neighborhood of
%the flow domain,  where $U'\neq 0$, or in a neighborhood of
%any  other point in the complex $y$-plane where this is so,
%this map is conformal.  Hence on these neighborhoods the
%differential  equation thus obtained contains no  singularities
%except for the critical point, $u=c$. A power series expansion
%about any ordinary point of the neighborhood gives an
%expression that converges uniformly on a neighborhood of fixed
%$u\neq c$ and $c$ otherwise arbitrary. An expansion about the
%critical point gives, upon transforming back to $y$,
%Eqs.~(\ref{2.6}) and (\ref{2.7}). (See also \citenum{RS66}.)

The Frobenius solutions can be used to construct neutral eigenmodes
(with $c$ real). We will go about this construction shortly;
however, Rayleigh's equation may also have complex eigenmode solutions,
for which $c$ is complex, and we  consider these first.

\subsection{Complex eigenmodes}
\label{com.emod}

When $c$ is complex, there is no singular point in Rayleigh's
equation for $y\in [-1,1]$, and we may then eliminate $\om$ between
(\ref{2.3}) and (\ref{2.4}) to obtain an integral equation for the
streamfunction (eigenfunction) of the complex eigenmodes:
\bq
\psi(y,c) = \int_{-1}^1 \cag(y,y') {U''(y')\psi(y',c)\over U(y')-c}dy'
.
\label{2.8}
\eq
This has the form of a Fredholm equation, and if there is a solution
for $c=c_D$, that solution is known to be
unique ({\it e.g.}\  \citenum{T85}).
Moreover, the conjugate, $c=c_D^*$, is also a solution.
These complex conjugates comprise a pair of discrete
eigenvalues, of which there can only be a finite number.
In fact, at the end of this paper, we essentially give an upper bound
on their number.

The complex pairs that satisfy (\ref{2.8}) are equivalent
to the homogeneous solutions of (\ref{2.5b}). Alternatively,
the values $c=c_D$ and $c=c_D^*$ are the
zeros of the Wronskian  (\ref{2.5f}) of the Laplace transform theory.

According to the Rayleigh-Fj{\o}rtoft Theorem, instabilities
occur only if the velocity profile contains a minimum in vorticity
(since we assume $U'>0$, the vorticity is everywhere
{\sl negative}). Thus, because ${\rm Im\ } c > 0$ signifies instability, the
theorem implies that the complex, discrete eigenmodes can only exist if
the vorticity has a minimum.

\subsection{Neutral  discrete eigenmodes}
\label{neu.emod}

The solution of Rayleigh's problem for the neutral eigenmodes
is not so clear cut. In fact, if $c$ is real, then there
can be no nontrivial, regular solutions for the
streamfunction, with only a single exceptional type of mode.
This follows from two facts. First, if $c<U(-1)=U_1$ or $c>U(1)=U_2$,
then Rayleigh's equation with the assumed
boundary conditions has only a trivial solution.
Thus neutral solutions must have critical layers.
Second, if $c$ lies within the flow
domain, $[U_1,U_2]$, then it can be shown that $\psi_g$ cannot
satisfy the boundary conditions ({\it e.g.} \citenum{B64}).
Therefore, all neutral eigenmodes must contain the
``bad'' solution $\psi_b(y,c)$, in which case
$\psi(y,c)$ must have an undefined derivative.
This failure to construct real analytic neutral
eigenfunctions is connected to the presence of a continuous spectrum
that occurs for wavespeeds lying inside the flow domain:
$c \in [U(-1),U(1)]$ \citenum{C60}.

The exceptional class of modes are those for which the critical layer
lines up with an inflexion point of the equilibrium flow profile.
In this case, $U''(y_c)=0$, and the singular
point is removed from Rayleigh's equation. The Frobenius solution,
$\psi_b(y,c)$, is then an analytic function.
The exceptional modes also satisfy
equation (\ref{2.8}), but now $c=U(y_c)=U(y_I)$, where
$y_I$ is the location of an  inflection point.
We call these eigensolutions, ``inflection-point modes.''
These are discrete eigenmodes embedded in the continuous spectrum.

One important feature of the inflection-point modes is that
they must be the limits of the complex pairs as ${\rm Im}(c)\rightarrow0$.
In fact, they are the basis of the Tollmien-Lin perturbation theory,
which builds the nearby complex solutions from the inflection-point mode.
This perturbation expansion is normally performed with
an underlying assumption that the limit exists. Though this is largely
a technicality, the assumption is strictly only valid when
$U(y)$ is an analytic function on $[-1,1]$
(this is one of the results that comes directly out of the complex analysis
described here).

\subsection{Singular eigensolutions}
\label{neu.seigs}

Neutral eigenmodes
cannot, therefore, be real analytic at the critical layer if
$U''(y_c)\ne0$. Consequently,
because the derivative of the streamfunction is not defined
for these modes, we can only impose continuity on such
solutions to Rayleigh's equation. Thus $\psi(y,c)$ may have
an arbitrary jump in its first derivative at $y=y_c$.
Any family of such solutions can be taken to be
a set of singular eigenfunctions of the continuous
spectrum. However, there is one family that is especially
useful, and we now construct these special singular eigenmodes.

The singular eigenfunctions are generalized
function solutions to Rayleigh's equation (see \citenum{VK55} and
\citenum{BM96}):
\bq
\ome(y,c) = \calp {U'' \psi\over U-c} + \lambda(c)\,
\delta[U(y)-c] , \label{2.9}
\eq
where $\calp$ indicates the Cauchy Principal Value, $\delta(U-c)$
is Dirac's delta function, and $\lambda$
is yet to be determined. With the interpretation of the
singularity in (\ref{2.9}) by means of the Cauchy principal value,
we may define the jump in the derivative of $\psi$ by $\lambda/U'(y_c)$.
This quantity is currently arbitrary and in particular
can be taken to be any function of $c$. Our special singular
eigenmodes arise from a judicious choice for  $\lambda$.

If we integrate
(\ref{2.9}) across the channel, we obtain
\bq
\Xi(c) :=
\int_{-1}^1 \ome(y,c) dy =
\calp \int_{-1}^1 {U''(y)\psi(y,c) \over U(y)-c}dy
+ {\lambda\over U_c'} ,
\label{2.10}
\eq
or
\bq
{\lambda\over U_c'}
= \int_{-1}^1 \ome(y,c) dy
- \calp \int_{-1}^1 {U''(y)\psi(y,c) \over U(y)-c}dy
\label{2.11}
\eq
\bq
=: \Xi(c) - \calp \int_{-1}^1 {U''(y)\psi(y,c) \over U(y)-c}dy ,
\label{2.12}
\eq
where $U_c'=U'(y_c)$ and $\Xi(c)$ is a parameter that
is, in general, a function of the eigenvalue. In fact,
since $\omega$ is a generalized function, the amplitude of
the eigenmode can only be fixed on multiplying by a
suitable test function and integrating. In
Eq.~(\ref{2.11}), the test function is simply unity, and
so $\Xi$ can be regarded as the eigenfunction amplitude.
Moreover, since this is a linear problem, we may
choose the dependence of $\Xi$ as we wish; this then amounts to
the selection of $\lambda$.

\def\cad{{\cal D}}

On using (\ref{2.4}), (\ref{2.9}), and (\ref{2.12}), we find:
\bq
\psi(y,c) =
{\lambda\over U_c'} \cag(y,y_c) + \calp \int_{-1}^1 \cag(y,y')
{U''(y')\psi(y',c)\over U(y')-c}dy'
\label{2.13}
\eq
\bq
= \Xi(c) \cag(y,y_c) +
\int_{-1}^1{ \cag(y,y')-\cag(y,y_c) \over U(y')-c}U''(y')
\psi(y',c)dy'
.
\label{2.14}
\eq
Although (\ref{2.13}) is at first sight  a
singular integral equation, (\ref{2.14}) is a regular Fredholm
equation of the second kind ({\it e.g.} \citenum{T85})
that is straightforward to solve.

Provided the Fredholm equation
has no homogeneous solution, (\ref{2.14})
has a unique particular solution. If (\ref{2.14}) does have a
homogeneous solution, particular solutions are, in general,
unbounded. The important point is
that a family of well-behaved singular eigenmodes
is needed to represent  the continuous spectrum.
If there are no homogeneous solutions, any choice for $\Xi(c)$
will suffice for this purpose. But if there are homogeneous solutions,
 a more specialized choice must be made. One possible selection is
$\Xi(c)=\cad(c)$, where $\cad(c)$ is the Fredholm
determinant. With this selection, the inhomogeneous term automatically
vanishes if there is a homogeneous solution. This ensures that
the solution to the Fredholm problem is always bounded.
Hence, even if there are homogeneous solutions, we can always
find a set of sensible singular eigenmodes.
(Note that $\cad(y_c)$ is determined by the kernel in (\ref{2.14}),
that is $\cag(y,y')$ and $U(y)$, and is independent of the
solution, $\psi(y,c)$, and so there is freedom for this
selection for $\Xi(c)$.)

For Rayleigh's problem with the profiles we have considered,
we have not found any homogeneous solutions to (\ref{2.14}).
So $\Xi(c)=1$ is a convenient choice in any practical application.
Then, from (\ref{2.9}) and (\ref{2.12}), we compute the
singular eigenfunction and the jump in the streamfunction's
derivative.

Finally, note that the amplitude of the solution at the
critical layer, $\psi(y_c,c)$, cannot vanish, since the Frobenius solution
(\ref{2.7}) satisfies $\psi_b(y_c,c)\neq 0$. This is an important property of
the singular eigenfunctions that will be used later.

%the Frobenius series solution
%(\ref{2.7}) can be shown to satisfy
%$\psi_b(y_c,c)=\varphi_3(y_c,c)={\rm constant}$.

%The solution to the Fredholm problem (\ref{2.14}) can be
%formally written in terms of the resolvent kernel,
%$\car(y,y';y_c)$:
%\bq
%\psi(y,y_c) = \cag(y,y_c) + \int_{-1}^1 \car(y,y';y_c)
%\cag(y',y_c) dy' .
%\label{2.15}
%\eq
%{}From this expression it is clear that properties such as
%continuity, differentiability, and so on, of $\psi$ follow
%from the dependence of $\car$ and $\cag$ upon $y$ and $y_c$.
%In turn, the nature of the dependence of $\car$ on $y$ and
%$y_c$ follows from the uniform convergence of the sequence
%of functions that approach $\car$, in the usual calculations
%of the resolvent ({\it e.g.}\ \citenum{T85}). This reveals
%that $\psi(y,y_c)$ is a real analytic function of $y$,
%except at $y=y_c$, where its derivative is discontinuous.
%Instead of pursuing this tack, we investigate the analytic
%nature of the solution to (\ref{2.14}), which will be used
%below and in Section~\ref{nyq}, by returning to Rayleigh's
%differential equation and enforcing a continuity condition at the
%critical layer.

%%%%%%%%%%%%%%%%%%%%%%%%%%%%%%%%%%%

\section{The dispersion relation}
\label{disp.rel}

In this section we construct an expression for the dispersion
relation. The form of the dispersion
relation is designed to facilitate the subsequent Nyquist analysis,
and is one that is useful for relating discrete eigenmodes to
continuum eigenmodes.

\subsection{Discontinuity and the dispersion relation}
\label{discont}

We begin by considering an arbitrary point $y_*$
that is  not coincident with the critical layer;
{\it i.e.} $y_*$ is any ordinary point of the differential equation.
For convenience we choose $y_*\in[-1,1]$ and we assume that $c$ is fixed in the
upper half $c$-plane.

Now we construct two solutions, $\Psi_<(y,c)$ and $\Psi_>(y,c)$,
that are defined for $y\in [-1,1]$
on the complex $c$-plane. The first,
$\Psi_<(y,c)$,  is defined by series expansion about the left
boundary point $y=-1$  and satisfies the
boundary condition $\Psi_<(-1,c)=0$.
This is a one-parameter family of solutions,
where the parameter can be taken to be a multiplicative
constant; {\it i.e.}\ we can write $\Psi_<(y,c)=C_<\Phi_<(y,c)$,
where $\Phi_<(y,c)$ is a parameter-independent solution of
Rayleigh's equation that satisfies $\Phi_<(-1,c)=0$  and
$\Phi_<'(-1,c)=1$.  Similarly, a second
one-parameter family of solutions is constructed by series
expansion about the point $y=1$. We
denote this second solution by  $\Psi_>(y,c)=C_>\Phi_>(y,c)$,
where by construction $\Phi_>(1,c)=0$ and $\Phi_>'(1,c)=1$. We
emphasize that at fixed  $y\in [-1,1]$,
both $\Phi_<(y,c)$ and $\Phi_>(y,c)$ are analytic in $c$ for all
$c\neq U(y)$.

The next step in our construction is to choose the
constants $C_<$ and $C_>$  so that $\Psi_<$ and $\Psi_>$ are
continuous at the point $y_*$; {\it i.e.}\ we set
$\Psi_<(y_*,c)=\Psi_>(y_*,c)$. This requirement leaves a remaining
constant that is an overall scaling factor; the two solutions can
be represented as follows:
\bqy
\Psi_<(y,c,y_*)&=& C(c,y_*)\Phi_>(y_*,c)\Phi_<(y,c)
\nonumber\\
\Psi_>(y,c,y_*)&=& C(c,y_*)\Phi_<(y_*,c)\Phi_>(y,c) \,,
\label{D.1}
\eqy
where the sole remaining constant is $C(c,y_*)$, in which we
have included dependence upon $c$ and $y_*$. Clearly, we are free to
choose $C(c,y_*)$ arbitrarily. We give a prescription for defining
$C(c,y_*)$ shortly.

No matter which (nonzero) value is chosen for $C(c,y_*)$, the derivatives
of the functions  $\Psi_<$ and $\Psi_>$ will not, for  general
values of $c$, match at the point $y_*$. However, in the event
that they do match, the value of $c$ is an eigenvalue and the
functions  $\Psi_<$ and $\Psi_>$ define an eigenfunction.  In
fact, when this is the case, $\Psi_<$ and $\Psi_>$ are analytic
continuations of each other. This follows because $y_*$ is an
ordinary point and both $\Psi_<$ and $\Psi_>$ are solutions of the
Cauchy problem for Rayleigh's equation with identical
specification of their values and derivatives at $y_*$. Moreover,
when the derivatives of the solutions match, the point  $y_*$ is
immaterial and the dependence upon $y_*$ drops out of the
incipient eigenfunction.

In light of the above, the jump in the derivatives of $\Psi_<$ and
$\Psi_>$ at $y_*$ is equivalent to a dispersion relation: its
vanishing determines $c=c(k)$. (Recall that the $k$ dependence has
been suppressed in the expressions above\@.)  Note that if $C(c,y_*)$
is chosen to be an analytic function of  $c$ in the upper half
plane, then $\Psi_<(y_*,c)$ and $\Psi_>(y_*,c)$ are also
analytic functions of $c$ in the upper half plane, which follows
from the assumption $U(y_*)\neq c$. Hence, the dispersion
relation is an analytic function for $c$ in the upper half
plane. Also, note that the dispersion relation  does not depend
upon the point
$y_*$, since matching of the derivatives at any point will give an
eigenfunction.

\subsection{Relationship between $\epsilon$ and $W$}

We now obtain a convenient expression for the dispersion relation.

In general, for $y$ within  the flow domain, we write the streamfunction
in the form,
\bq
\Psi(y,c,y_*):=H(y-y_*)\Psi_>(y,c,y_*) +
H(y_*-y)\Psi_<(y,c,y_*)
\label{D.1b}
\eq
and  $H$ is the Heaviside function. The vorticity,
by which we mean $\om(y,c,y_*)=\Psi'' -k^2\Psi$,
then has a representation,
\bq
\om(y,c,y_*)= \Om(y,c,y_*) + \tilde\epsilon(c,y_*)\, \delta(y-y_*)\,.
\label{D.2}
\eq
where $\tilde\epsilon(c,y_*)$ measures the jump
in the derivatives,
\bq
\tilde\epsilon(c,y_*):=\Psi_>'(y_*,c,y_*)-\Psi_<'(y_*,c,y_*)\,,
\label{D.3}
\eq
and
%$\Om(y,c,y_*)$ is analytic for $y$ in a neighborhood that
%contains  the flow domain. In fact, since  $\Psi_<$ and $\Psi_>$ are
%solutions of Rayleigh's equation,
\bq
\Om(y,c,y_*) = \left\{
\matrix{
U''(y)\Psi_<(y,c,y_*)/[U(y)-c]
& {\rm if \ \ } y> y_* \cr
U''(y)\Psi_>(y,c,y_*)/[U(y)-c]
& {\rm if \ \ } y< y_*\,.
}
\right.
\label{D.4}
\eq
By inserting (\ref{D.1}) into (\ref{D.3}) we see that
\bqy
\tilde\epsilon(c,y_*)
&=&C(c,y_*)\left[\Phi_>(y_*,c)\Phi_<'(y_*,c)
- \Phi_<(y_*,c)\Phi_>'(y_*,c)\right]
\nonumber\\
&=&:C(c,y_*)W(c)\,,
\label{D.4b}
\eqy
where $W$, the Wronskian for Rayleigh's equation, is
independent of $y_*$.
%Note that (\ref{D.4b}) implies
%that $\epsilon(c)$ does not depend on $y_*$ provided
%we choose $C(c)$ independently of $y_*$.
Note that this quantity can be identified with the Wronskian of
(\ref{2.5e}) by substituting $p=-ikc$ into the latter.

%%%%%%%%%%%%%%%%%%%%%%%%%%%%%%%%%%%

We next normalize the solution by integrating (\ref{D.2}) over the
flow domain and setting the result equal to the  $c-$dependent
parameter, $\tilde\Xi(c,y_*)$:
\bq
\int_{-1}^{+1}\om(y,c,y_*)dy=\tilde\Xi(c,y_*)\,,
 \label{D.5}
\eq
which yields
\bq
\tilde\Xi(c,y_*) =
\tilde\epsilon(c,y_*) +
\int_{-1}^{y_*}\frac{U''(y)\Psi_<(y,c,y_*)}{U(y)-c}\, dy +
\int_{y_*}^{1}\frac{U''(y)\Psi_<(y,c,y_*)}{U(y)-c}\,dy
.
\label{D.6a}
\eq
By using Rayleigh's equation, we then find
\bq
\tilde\Xi =
  -\Psi_<'(-1,c,y_*) + \Psi_>'(1,c,y_*)
 - k^2 \int_{-1}^{y_*}\Psi_<(y,c,y_*) \, dy
-  k^2 \int_{y_*}^{1}\Psi_>(y,c,y_*)\,dy
\,.
\label{D.6}
\eq
%\bqy
%\Xi(c)&=&\epsilon(c) + \int_{-1}^{1}\Om(y,c,y_*) dy
%\nonumber\\
%&=&\epsilon(c)
%+ \int_{-1}^{y_*}\frac{U''(y)\Psi_<(y,c,y_*)}{U(y)-c}\, dy +
%\int_{y_*}^{1}\frac{U''(y)\Psi_<(y,c,y_*)}{U(y)-c}\,dy
%\nonumber\\
%&=& \epsilon(c)
%+ \int_{-1}^{y_*}\left[\Psi_<''(y,c,y_*) -  k^2\Psi_<(y,c,y_*)\right] dy
%+
%\int_{y_*}^{1}\left[\Psi_>''(y,c,y_*) -  k^2\Psi_>(y,c,y_*)\right] dy
%\nonumber\\
%&=&  -\Psi_<'(-1,c,y_*) + \Psi_>'(1,c,y_*)
% - k^2 \int_{-1}^{y_*}\Psi_<(y,c,y_*) \, dy
%-  k^2 \int_{y_*}^{1}\Psi_>(y,c,y_*)\,dy
%\,.
%\label{D.6}
%\eqy
%In the penultimate equality of (\ref{D.6}) Rayleigh's equation has been
%used, while in the last equality of (\ref{D.6}) the definition of
%$\eps$ as given in  (\ref{D.3}) has been used.
Finally, by inserting
(\ref{D.1}) into (\ref{D.6}) we may express $C(c,y_*)$ in terms of
$\tilde\Xi(c,y_*)$ (or {\it vice versa}):
\bqy
C(c,y_*)=-\tilde\Xi(c,y_*) \Big[
k^2 \int_{-1}^{y_*} \Phi_>(y_*,c)\Phi_<(y,c)\, dy
&+&
k^2 \int_{y_*}^{1} \Phi_<(y_*,c)\Phi_>(y,c)\,dy
\nonumber\\
&+& \Phi_>(y_*,c) -\Phi_<(y_*,c)
\Big]^{-1}
\,.
\label{D.6b}
\eqy
Because $\Phi_<$ and $\Phi_>$ are analytic in $c$ in the upper half plane
for fixed $y$, the quantity in the denominator of the right-hand side
of (\ref{D.6b}) is also analytic.
Thus, if we were to select $C(c,y_*)$ to be analytic in the upper half
$c-$plane, then $\tilde\Xi(c,y_*)$ would also be. However,
the converse is not quite true:
if $\tilde\Xi(c,y_*)$ is selected to be analytic in the upper half plane,
then $C(c,y_*)$ is also analytic except, perhaps, for
poles at the zeros of the denominator. We will interpret these zeros shortly.

\iffalse
Now given the $\Psi$ of (\ref{D.1b}), we can use the second
equality of (\ref{D.6}) to finally write an expression for the {\it
dispersion relation}:
\bq
\epsilon(c)= 1 -
 \int_{-1}^{1}\frac{U''(y)\Psi(y,c,y_*)}{U(y)-c}\, dy \,.
\label{D.9}
\eq
In this expression we have not included dependence upon
$y_*$ since the  vanishing of $\ep$ gives an eigenvalue independently of
the matching point. In summary, $\epsilon(c)$, by its construction, is an
analytic function for $c$ in the upper half plane that vanishes when $c$
is an eigenvalue there.  Its analyticity follows from that of $\Psi$
and the well-known  property of Cauchy type integrals.  We also note that
$\ep(c)$ possesses a branch cut that extends along the flow domain,
$[-1,1]$.

\fi
%%%%%%%%%%%%%%%%%%%%%%%%%%%%%%%%%%%

%Note that,
%since Rayleigh's equation has real coefficients,  $\epsilon^*(c)$,
%the complex conjugate of  $\epsilon(c)$, is analytic for $c$ in the
%lower half plane, and its vanishing gives the eigenvalues that are
%the complex conjugates of those determined by $\epsilon(c)$ in the
%upper half plane.

\subsection{Rayleigh Green function and singular eigenfunctions}

The next piece of the puzzle is to relate the
dispersion relation to the singular eigenfunctions.

%Since the above construction was
%somewhat involved,  we  now give  and alternative interpretation of
%what was done.

We may rewrite (\ref{D.2}) in the form,
\bq
\Psi'' -k^2 \Psi -\frac{U''\Psi}{U-c}=\tilde\ep\,  \delta(y-y_*)\,.
\label{D.10}
\eq
Thus, $\Psi$ is related to the Green function for Rayleigh's
equation. This equation is a bit subtle, since $\tilde\eps$
(according to (\ref{D.6a})) is in fact a
property of the solution and so the equation is self-referential.
However, if  $y_*\neq y_c$ and  $\tilde\ep\neq 0$,
then $\Psi/\tilde\ep$ satisfies the normal
equation for the Green function.

We may turn equation (\ref{D.10}) into one of integral form
on using the Green function of the Laplacian:
\bq
\Psi(y,c,y_*)=
\calg(y,y_*)\tilde\epsilon(c,y_*) +
\int_{-1}^{1}\calg(y,y')\frac{U''(y)\Psi(y',c,y_*)}{U(y')-c}\,
dy'\,.
\label{D.13}
\eq
Moreover, on using the relation (\ref{D.6a}), this can be written as
\bq
\Psi(y,c,y_*)= \tilde\Xi(c,y_*) \calg(y,y_*) + \int_{-1}^1 \Psi(y',c,y_*)\,
\calk(y,y',y_*,c)\, dy'\,,
\label{D.11}
\eq
where
\bq
\calk(y,y',y_*,c):=U''(y')\frac{\left[\calg(y,y')-\calg(y,y_*)\right]}
{U(y')-c}\,.
\label{D.12}
\eq
Hence the Green function can be constructed by solving another
Fredholm integral equation.

In fact, the two Fredholm problems are
closely related: if we take the limit $c\rightarrow c_r+i0\equiv U(y_c)+i0$
with $y_*=y_c$, then we recover the Fredholm
equation for the singular eigenfunctions from (\ref{D.11})--(\ref{D.12})
with $\Psi(y,c_r+i0,y_c)=\psi(y,c_r)$ and
$\tilde\Xi(c,y_c)\equiv\Xi(c)$.

Similarly, we may recover the singular eigenfunction (\ref{2.9})
from the Green function equation (\ref{D.2}) in the same limit:
we let $c\rightarrow c_r+i0\equiv U(y_c)+i0$
in (\ref{D.6a}), giving
\bq
\tilde\epsilon(c_r+i0,y_c) =
\tilde\Xi(c_r+i0,y_c) -
\int_{-1}^{1}\frac{U''(y)\Psi(y,c_r+i0,y_c)}{U(y)-c_r-i0}\, dy
.
\label{D.6ar}
\eq
However, since $\Psi(y,c,y_*)$ is analytic in the upper half plane,
there exists a generalized form of the Plemelj relation
({\it e.g.} \citenum{G90}), which leads to
\bq
\tilde\epsilon(c_r+i0,y_c) = \Xi(c_r) -
\calp \int_{-1}^{1}\frac{U''(y)\psi(y,c_r)}{U(y)-c_r}\, dy
- i \pi {U''(y_c)\psi(y_c,c_r) \over U'(y_c)}
\label{D.6ar2}
\eq
\bq
=: \tilde\epsilon_r(c_r,y_c)+i\tilde\epsilon_i(c_r,y_c)
,
\eq
on using the association of the Fredholm equations
to replace $\Psi(y,c_r+i0,y_c)$ with $\psi(y,c_r)$.
The Plemelj relation also implies that
\bq
\omega(y,c_r+i0,y_c)
= \calp {U''\psi(y,c_r)\over U-c_r} + \tilde\ep_r(c_r,y_c) \delta(y-y_c)
.
\eq
Finally, we identify $\omega(y,c_r+i0,y_c)$ with $\omega(y,c_r)$,
$\tilde\epsilon(c_r,y_c)$ with
$\epsilon(c_r)$, and $\tilde\epsilon_r(c_r,y_c)$ with $\lambda(c)/U_c'$.
With these associations understood, we will drop the tildes
in the next section and the extra argument in $y_c$.

Note that we could take a completely different approach to the problem
beginning from the Green function.
First we select $\tilde\Xi(c,y_*)$ to be a function that
is suitably analytic in $c$. Then,
the solution of the integral equation
(\ref{D.11}) can be shown to have the various analyticity
properties we have found for $\Psi(y,c,y_*)$ by using
Fredholm theory (the kernel is an analytic function
of the parameter  $c$  in the upper half plane).
{}From there, we build the quantity $\tilde\epsilon(c,y_*)$,
which we know has zeros if $c$ is an eigenvalue.
Moreover, equation (\ref{D.6b}) determines a function
$C(c,y_*)$ by which we may relate the solution of the
integral problem to the solutions $\Phi_<$ and $\Phi_>$
used above.

The only fly in the ointment is the possibility of a homogeneous solution
to the Fredholm problem for some value of $c$, in which case
the particular solution we seek is unbounded. {}From Fredholm theory
we know that the solution
$\Psi$ is not analytic at this value of $c$, and, in fact, has a pole.

However, this nonanalyticity is evidently an artifact of
solving the Fredholm problem, since the solutions $\Phi_<$ and $\Phi_>$
are analytic in $c$. Moreover, since they are related to the
solution $\Psi$ of (\ref{D.11}) simply by the factor $C(c,y_*)$
as in equation (\ref{D.1}), it is clear
that the nonanalyticity in $\Psi(y,c,y_*)$ is equivalent to
a pole in $C(c,y_*)$. But since we may choose $\tilde\Xi(c,y_*)$
to be analytic, the pole must arise from a zero in the denominator
of (\ref{D.6b}). This is the advertized interpretation of the
poles in $C(c,y_*)$; they correspond to the presence of a
homogeneous solution to the Fredholm problem (\ref{D.11}).

At this stage, two remarks are in order. We have already noted
that there are no homogeneous solutions for $c$ in the flow domain
in the context of the singular eigenfunction equation.
Hence, $C(c,y_*)$ must be real and contain no poles for these
values of $c$. Secondly, since this nonanalyticity is $\Psi(y,c,y_*)$
is purely an artifact of solving the Fredholm problem (\ref{D.11}),
we can take a slightly different tack and make a judicous
choice for $\tilde\Xi(c,y_*)$ that avoids the problem.
Again, this is just a choice like $\tilde\Xi=\cald(c,y_*)$, where
$\cald(c,y_*)$ is the Fredholm determinent of the kernel in
(\ref{D.11}). Once we make this choice, we may derive a
solution of the integral equation that is analytic in $c$
in the upper half plane, and use it to build the
dispersion relation through $\tilde\epsilon(c,y_*)$.
However, we will explicitly follow the route outlined earlier in this
section rather than this Green-function based avenue.

\section{Nyquist method}
\label{nyq}

In the previous section we discussed the functions  $\epsilon(c)$ and
$W(c)$,  and their relationship.  If their exists a discrete value of the phase
velocity, $c=c_D$, such that $W(c_D)=0$, then we have an exponentially growing
eigenmode with growth rate, ${\rm Im}(kc_D)$. By construction  we know that
$W(c)$ is analytic for $c$ in the upper half plane, and that it also has a
branch cut along the real axis between $U_1:=U(-1)$ and
$U_2:=U(+1)$. This branch cut arises from that of the natural logarithm of the
Frobenius solution  and the matching procedure ({\it cf.}\  (\ref{W.chi})
of the
Appendix.) Because of these properties we begin with
$W$ in our analysis below, but in the end we express the final result in
terms of
$\epsilon$, a quantity that is  by design  reminiscent of the dispersion
relation of Vlasov theory.

The Nyquist method relies upon the argument principle of complex analysis.
In the present context this principle implies that the integral,
\bq
 {1\over2\pi i} \int_\cac {W'(c)\over W(c)}\, dc
\label{3.2}
\eq
counts the number of zeros of $W$  in the region enclosed by a closed
contour $\cac$  in the $c-$plane.  We choose  $\cac$  to run along the $c_r$
axis, with $c_i$ fixed to an arbitrarily 
small positive value, and then the contour is
closed by a large  semicircular portion as shown in Fig.~\ref{fig:nf2}. 
As the radius of the  semicircle  goes to infinity,  (\ref{3.2}) gives 
the number of exponentially growing eigenvalues. 
Equivalently, (\ref{3.2}) is the number of times the path
determined by the function $W(c)$ encircles the origin of the
$(W_r,W_i)$ plane as $c$ completes a circuit of $\cac$.

\begin{figure}
\begin{center}
\leavevmode
\epsfysize=5.5cm
\epsfbox{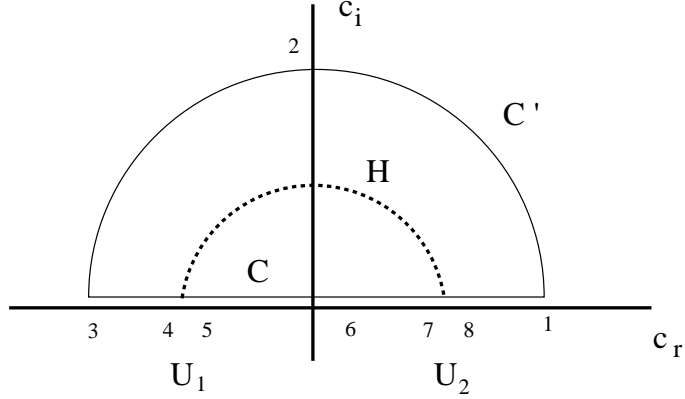}
\end{center}
\caption{The contours $\calc$, $\calc'$, and $H$ in the $c$-plane. The
closed contour $\calc$  runs the entire circuit  from $1\rightarrow
2\rightarrow\dots \rightarrow 8\rightarrow 1$, with the portion from
$3\rightarrow1$ lifted  infinitesimally above the real axis. The contour
$\calc'$ is   $\calc$ with the piece along the flow domain,
$5\rightarrow7$,  removed. The contour $H$ (dashed) is Howard's semicircle,
within which the unstable eigenvalues must lie.}
\label{fig:nf2}
\end{figure}

It is efficacious to decompose the integral (\ref{3.2}) into the following two
pieces:
\bq
 {1\over2\pi i} \int_\cac {W'(c)\over W(c)}\, dc  = {1\over2\pi i}
\int_{U_1+i0}^{U_2+i0} {W'(c)\over W(c)} dc + {1\over2\pi i} \int_{\cac'}
{W'(c)\over W(c)} dc.
\label{3.3}
\eq
The first integral is all important, while the second is relatively minor.
To understand this recall  Howard's semicircle theorem \citenum{DH66}, which
states that the zeros of $W$ must lie within a disk of radius,
$(U_1+U_2)/2$, centered at the point $(U_2-U_1)/2$ (as illustrated in
Fig.~\ref{fig:nf2}). Hence, if $\cac$ encloses the semicircle, it contains
all of the unstable eigenvales. In fact, wherever  $W$ is analytic, we may deform the
contour $\cac'$.  Since  $W$ is analytic in the upper half plane we may
deform $\cac'$ into any contour there that connects  
$W(U_1+i0)$ to $W(U_2+i0)$ 
%(again indicated by the dashed semicircle, $H$,  of
%Fig.~\ref{fig:nf2}). 
The important point is that  since $\cac'$ is deformable to
any other contour lying outside Howard's semicircle, the path defined by $W(c)$
as $c$ varies along $\cac'$ cannot lead to a new enclosure of the origin
because this would mean a zero of $W$ outside the semicircle. In other words, the
count of unstable eigenvales must be independent of the integration around $\cac'$; the
only importance of the integral over $\cac'$ is to complete a closed path
in the $(W_r,W_i)$ plane without encircling the origin. In the Appendix we
demonstrate this explicitly by extending the contour $\cac'$ to
infinity and then by analyzing the image of its various pieces in the
$W$-plane.  The upshot is that we may ignore the
$\cac'$  part of the contour:  the change in the argument of W along the flow
domain is equal to the number of times the function $W(c)$ encircles the origin, which is equal to the number of
unstable eigenvalues.

In Section~\ref{disp.rel} we showed that  $\epsilon$ and $W$ are related by
$\epsilon(c)= C(c)\,W(c)$. But as $c$
traverses the flow domain:
\bq
\int_{U_1+i0}^{U_2+i0}\, \frac{\epsilon'}{\epsilon}\, dc
= \int_{U_1+i0}^{U_2+i0}\, \frac{W'}{W}\, dc
+ \int_{U_1+i0}^{U_2+i0}\, \frac{C'}{C}\, dc\,.
\label{FD.int}
\eq
Upon introducing
\bq
\epsilon = |\epsilon|\,e^{i\,{\rm arg} [\epsilon]}\,  \quad
{\rm and} \quad
W = |W|\,e^{i\,{\rm arg} [W]}\,
,
\eq
(\ref{FD.int}) becomes
\bq
 \Big(\ln|\epsilon| + i\,{\rm arg} [\epsilon]\Big)\,\Big|_{U_1+i0}^{U_2+i0}
=\Big(\ln|W| + \ln C + i\,{\rm arg} [W] \Big)\,\Big|_{U_1+i0}^{U_2+i0}\,.
\eq
Now, as remarked at the end of Section \ref{disp.rel}, $C(c)$ is real and
finite along this part of the contour. Hence,
equating imaginary parts gives $\Delta {\rm arg} [\epsilon]
= \Delta{\rm arg} [W]$.
%The point is that the argument of $C$ remains
%zero on the contour if $C$ is analytic in a neighborhood of the flow
%domain.

Therefore, the  number of unstable eigenvalues is  determined by the change in
the argument of the following quantity:
\bq
\epsilon(c_r+i0) =
1 - \calp \int_{-1}^1 {U'' (y) \psi(y,c)\over U(y)-U(y_c)}dy
- i\pi{U''(y_c)\psi(y_c,c)\over U'(y_c)}
\label{3.4}
\eq
\bq
=\epsilon_r(c_r)+i\epsilon_i(c_r)\,,
\label{3.5}
\eq
where $c_r=U(y_c)$ (and we have explicitly made the choice $\Xi(c)=1$),
which  is the ``Nyquist function'' of Eq.~(2).
As $c_r$ varies from $U_1$ to $U_2$, or equivalently, $y_c$ from $-1$ to
1,  $\epsilon(c_r)$ executes a  {\it closed} path in the
$(\epsilon_r,\epsilon_i)$ plane. This path   begins and ends at $(1,0)$
since  $\psi(\pm 1,c)=0$.  In between, the path  circulates around and
the following considerations help us determine
whether or not the origin is encircled.

The path can only cross the $\epsilon_r-$axis at the
points for which $\epsilon_i=0$, or
$\psi(y_c,c_r)U''(y_c)=0$. However, as noted
at the end of Section \ref{form},
$\psi(y_c,c_r)$ cannot vanish. Hence
the crossing points are just the inflection points of the
velocity profile. We denote these by
$y_I$, with $U''(y_I)=0$ and $c=c_I=U(y_I)$.
Thus, the path emerges from the point $(1,0)$ on the
$(\epsilon_r,\epsilon_i)$ plane, circulates around crossing the
$\epsilon_r-$axis as many times as there are inflection points,
and then terminates at $(1,0)$. If the origin is encircled, we have
an exponentially growing instability. Hence, by calculating the
singular eigenfunctions from the Fredholm problem (\ref{2.14}) and
then constructing $\epsilon(c_r+i0)$, we can determine if there is an
unstable eigenvalue. 

In fact, the path  can only enclose the origin if
there is at least one crossing point
to the left of the origin. Such a crossing point is guaranteed if
\bq
\int_{-1}^1 {U''(y)\psi(y,c_I)\over U(y)-U(y_I)} dy > 1
\label{3.6}
\eq
for at least one of the inflection points.

Inequality (\ref{3.6}) is actually
a sufficient condition for instability. We see this as follows: 
if there are an odd number of crossings to the left of the
origin, then (\ref{3.6}) certainly guarantees an enclosure of
the origin. But if there are an even number, then one can
envision paths that cross to the left of the origin, but do
not enclose it. In this case, (\ref{3.6}) may still hold,
but the locus fails to encircle the origin. However,
the condition still predicts instability because
we may yet vary $k$, and the path must
change continuously  as we change $k$. {}From the Fredholm problem,
one can observe that $\psi\sim1/k$ for large $k$. This means
that $\psi\rightarrow0$ as $k\rightarrow\infty$. Thus for large
enough $k$, the path of $\epsilon$ shrinks about the point
$(1,0)$. Since varying $k$ also cannot destroy crossing points,
it must be the case that there is a range of values of $k$ for
which two of the crossing points must straddle the origin, and
the path encircles it. (An example exhibiting this feature is
given in Section~\ref{knot0}.)

Thus (\ref{3.6}) guarantees an enclosure of the origin for some value
of $k$, and for any number of crossings to the left of the origin.
In other words, we deduce the following
necessary and sufficient condition for instability:
\begin{quote}
{\it Rayleigh's equation possess an unstable eigenmode
if and only if there exists a wavenumber, $k$,  and  at least one
inflection point, $y_I$, such that Eq.~(\ref{3.6}) holds}.
\end{quote}

Note that, by ``crossing point'' we mean strictly only those
inflection points for which $U'''\ne0$. This implies that
intersection of the path  with the axis leads to a traversal of
the axis from one half plane to the other. Inflection points for
which $U'''=0$ can lead to the path  touching but not crossing
the axis, and these intersections will not signify a nearby
enclosure of the origin. However, an arbitrarily small deformation
of $U(y)$ can in this situation lead to the locus encircling the
origin. Thus profiles with $U'''=0$ correspond
to marginally stable states (see  Section~\ref{num}).

Unfortunately, the sufficient condition implied by (\ref{3.6})
has a more complicated dependence on $k$ than in the related
Vlasov problem (the $k-$dependence in the Vlasov problem is
contained purely in a multiplicative factor). The dependence here
comes from the dependence of $\psi(y,c_I)$ on $k$, which is not
obvious. Nevertheless, if $k^2 > k_m^2 := -{\rm Min\ }[U''/(U-c_I)]$
for all of the inflexion points $y_I$, then we can
apply a  result of Howard \citenum{H64} which indicates that there
can be no instability. Hence for $k> k_m$, the crossing point
must lie to the right of the origin, so we can at least bound the
range of interesting wavenumbers from above.

Finally, the condition (\ref{3.6}) refers only to
exponentially growing eigenmodes and, therefore,
does not encompass all the possibilities for instability.
It fails to account for a possible algebraic
instability of the continuous spectrum. However, in the
present context, we can safely ignore such a
possibility since we know that the continuum is stable for a
monotonic velocity profile \citenum{RS66}. Therefore,  algebraic instability
is ruled out. Hence (\ref{3.6}) is the encompassing condition for
instability. This may not be  true if the
profile is nonmonotonic (see \citenum{SR90}).

\section{Special cases}
\label{special}

In the previous section we described our necessary and sufficient condition
for instability; here we illustrate our criterion for two special
examples: general profiles with $k=0$, and an asymptotic
result.
%, where the analysis can be
%carried through. The Fredholm problem for the streamfunctions of the singular
%eigenfunctions can be solved explicitly in the case of
%the $k=0$ eigenmodes,
%while it can be solved approximately in the second example.

\subsection{General $k=0$ eigenmodes}

For an arbitrary profile with $k=0$, Rayleigh's equation simplifies and
the eigenmodes can be found by directly integrating.
It is straightforward to construct the Wronskian,
\bq
W(c) = (U_2-c)(c-U_1) \int_{-1}^1 {dy\over (U-c)^2}
.
\eq
We could analyze this dispersion relation independently of the machinery
developed in the previous sections. However, we consider the
$k=0$ problem by way of illustration, and so we apply the general
methodology.

The Fredholm problem (or the procedure of Section \ref{disp.rel}) for the
streamfunctions of the singular eigenfunctions has in this case the analytical
solution:
\bq
\psi(y,c) = \left\{ \matrix{
- (U_2-c)(c-U_1)(U_2-U_1)^{-1} U_c' (U-c)
\int_{-1}^y dy'/[U(y')-c]^{2} & {\rm for\ }
y < y_c \cr
(U_2-c)(c-U_1)(U_2-U_1)^{-1} U_c' (U-c)
\int_y^1 dy'/[U(y')-c]^2 & {\rm for\ }
y > y_c. \cr
}\right.
\label{4.3}
\eq
Consequently, the Nyquist function can be written in the
form,
\bq
\epsilon (c_r+i0) = -{(U_2-c_r)(c_r-U_1)\over U_2-U_1} U_c'
\int_{-1}^1 {dy\over [U(y)-c_r-i0]^2}
,\label{4.4}
\eq
or
\bq
\epsilon_r(c_r) = {(U_2-c_r)(c_r-U_1)\over U_2-U_1} U_c'
\left\{ {1\over (U_2-c_r)U_2'} + {1\over (c_r-U_1)U_1'}
+ \calp \int_{-1}^1 {U''(y)dy\over [U(y)-c_r]U'(y)^2} \right\}
\label{4.5}
\eq
and
\bq
\epsilon_i(c_r) =
- \pi {(U_2-c_r)(c_r-U_1) U''(y_c)\over (U_2-U_1) U'(y_c)^3} ,
\label{4.6}
\eq
where $U_1'=U'(-1)$ and $U_2'=U'(1)$.

Note that, for the $k=0$ problem, $C(c)=-U_c'/(U_2-U_1)$ 
for $c$ in the flow domain.
Thus $C(c)$ is indeed  real and finite, as we remarked earlier.

In the Nyquist construction, the important piece of the
contour $\cac$ leaves the point $(1,0)$, executes some path that
is dependent on the details of
$U(y)$, and finally returns to $(1,0)$. That is, the path is
closed,  as we remarked earlier. The Nyquist function of
(\ref{4.4}) is related to that obtained in \citenum{RS64}, where
this special case of $k=0$ was considered in a more specific
fashion. Note, however, that the  Nyquist function of
\citenum{RS64} differs from  (\ref{4.4}) by  a factor in front of the
integral that leads to
$\epsilon(c)$ vanishing at $c=U_1$ and
$U_2$. If this were not so, our Nyquist plots would not be closed
loops and one would be forced to consider the neglected piece of
the contour, $C'$, in detail.

\def\vare{{\varepsilon}}
\def\cau{{\cal U}}

\subsection{An asymptotic result}

Our second example is only an approximate result; it
concerns  velocity profiles of the form, $U(y) = y + \vare^2
\cau(y/\vare)$, where $\vare\ll1$. This kind of velocity profile
represents a linear background profile with a superposed, sharply
varying, ``defect.'' To leading order in $\vare$, the
corresponding streamfunction is given by
\bq
\psi(y,c) = \cag(y,y_c) + O(\vare)
,
\label{4.7}
\eq
and the Nyquist function has the simple form
\bq
\epsilon(c) =
1 - {\tanh k\over 2k} \int_{-\infty}^\infty
{\cau''(\eta)d\eta \over \eta - c}
+ O(\vare) ,
\label{4.8}
\eq
or
\bq
\epsilon_r(c_r) = 1 - {\tanh k\over 2k} \calp
\int_{-\infty}^\infty
{\cau''(\eta)d\eta \over \eta - c_r} + O(\vare)
\label{4.9}
\eq
and
\bq
\epsilon_i(c_r) = - {\tanh k\over 2k} \cau''(y_c) + O(\vare) .
\label{4.10}
\eq
This asymptotic result is given in \citenum{BDY96},
following \citenum{G65}. It
is closely related to the corresponding Vlasov solution.

Note that in the two examples,
$U(y)$ need not be analytic on $[-1,1]$; in fact, the
existence of two derivatives is sufficient.

\section{Numerical solutions for sample profiles}
 \label{num}

We now construct some Nyquist plots numerically for a
trio of sample profiles that illustrate different features of the
inviscid stability problem.

\subsection{Single inflection point}
\label{num.single}

The profile,
\bq
U(y) = \tanh \beta y
,
\label{4.11}
\eq
is an example of an equilibrium with a single inflection point.
Nyquist plots for various values of $k$ and $\beta=2$ are drawn in
Fig.~\ref{fig:nf1}(a). For wavenumbers $k<k_c$ with $k_c\simeq1.832$, there
is evidentally an unstable eigenvalue,  since the Nyquist plots
enclose the origin over this range of $k$.
Plots for $k=0$ and different values of $\beta$ are shown in
Fig.~\ref{fig:nf1}(b).

In Fig.~\ref{fig:nf1}(a), the Nyquist plot passes through the origin
without encircling it for the  critical value $k=k_c$. This  wavenumber is the
demarcation between stability and instability; that is, it is the stability
boundary,
$k_c=k_c(\beta)$. This stability boundary is displayed
in Fig~\ref{fig:nf5}(a). As $\beta$ decreases, the critical wavenumber
decreases until it vanishes at a special value of $\beta=\beta_m\simeq 1.2$.
For the profile with this critical parameter value,
the $k=0$ Nyquist plot passes through the origin and none encircle it. In
other words, such a profile is a marginally stable state.

\begin{figure}
\begin{center}
\leavevmode
\epsfysize=14.cm
\epsfbox{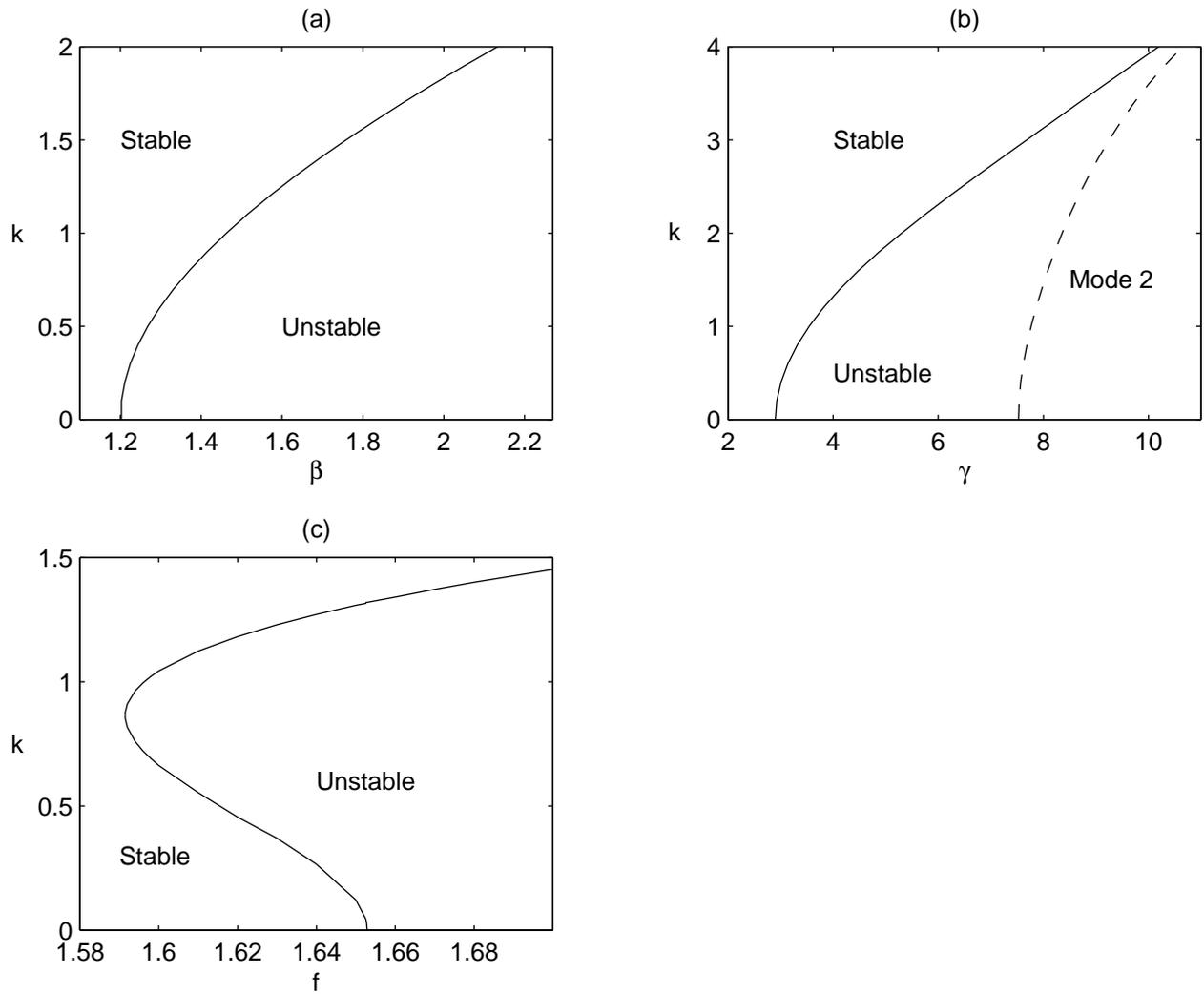}
\end{center}
\caption{Stability boundaries for the three equilibrium profiles of Sec.~7.
Figures (a) and (b) depict the nucleation of instability through $k=0$,
while (c) depicts the nucleation of instability through $k\neq 0$. 
In (b), the stability boundary of the second and third unstable modes
is also shown.}
\label{fig:nf5}
\end{figure}

This feature of the profile is an example of a general result:
instability always sets in first at $k=0$ for profiles with a single inflection
point  ({\it e.g.} \citenum{DH66}).  Hence the instability  condition is
given by $\epsilon_r<0$ with $\epsilon_r$ given by Eq.~(\ref{4.5}). This is
equivalent to the result of \citenum{RS64}.

In some applications, however, one may be interested in flows that
are periodic in $x$ (such as in annular or spherical geometry,
or in numerical simulations). In these cases, 
there is a minimum wavenumber, and $k=0$ is neither accessible
nor relevant. 
Hence the $k=0$ theory is not applicable even for a
single inflexion point. In this circumstance one must deal 
with the general Nyquist function and sufficient
stability condition described in Section 5.

\subsection{Multiple inflection points}

The profile,
\bq
U(y) = y + {1\over2\gamma}\sin \gamma y
,
\label{4.12}
\eq
is an example in which there are multiple inflection points. In this case,
the number of inflection points varies with $\gamma$.
For $\gamma=5$, $7.5$, and 10, there are 3, 5, and 7
inflecton points, respectively. 
As $\gamma$ increases, the new inflexion points appear in pairs
through the boundaries, $y=\pm1$.
The Nyquist plots at the three parameter
values and $k=1$ are shown in Fig.~\ref{fig:nf3}(a)--(c). For these plots there
are multiple encirclings of the origin, signifying
multiple instabilities. For example, when $k=1$ and $\gamma=10$,
there are three loops around the origin (see Fig.~\ref{fig:nf3}(c)).
As $\gamma$ increases and inflexion points appear at the boundaries,
the Nyquist curve acquires more loops that appear out of the 
asymptote $(1,0)$.

\begin{figure}
\begin{center}
\leavevmode
\epsfysize=14.cm
\epsfbox{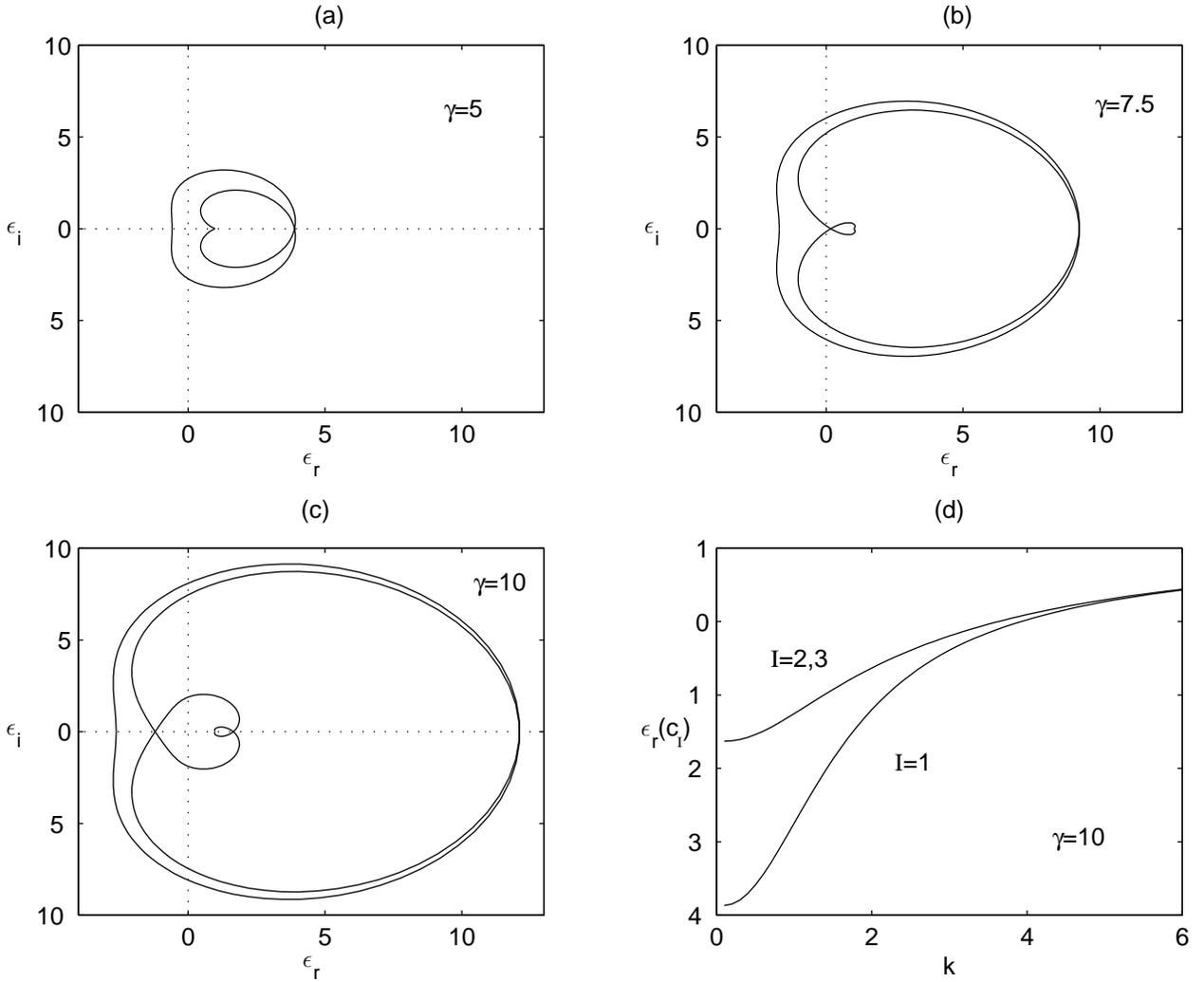}
\end{center}
\caption{Nyquist plots for the   multiple inflection point  profile,
$U(y)=y + (\sin\gamma y)/2\gamma$, with $k=1$. (a) The case $\gamma=5$  has
one encircling of the origin and thus one unstable eigenvalue. (b) In the case
with $\gamma=7.5$ two more loops have appeared and the
curve almost encircles the origin two more times. That is, it is
close to the threshold of the instability of two further modes.
(c) The case $\gamma=10$ has three encirclings
of the origin and three unstable eigenvalues. (d) Depiction of the monotonic
dependence of the crossing values to the left of the asymptote (1,0)  as a
function of $k$.}
\label{fig:nf3}
\end{figure}

At $\gamma=10$,  there are therefore three unstable eigenvalues with $k=1$,
which
is the most this profile can support (four of the inflection points lead
to crossing of the $\epsilon_r-$axis, but these lie to the
right of the asymptote $(1,0)$). Note that the antisymmetry
of the profile means that the path of  $\epsilon(c_r+i0)$ is
symmetric under reflection about the $\epsilon_r-$axis.

If we vary $k$, and calculate $\epsilon_r(c_I)$, where $c_I$ with 
$I=1,2,3$  denotes the three inflection points that lead to
crossings to the left of the asymptote (1,0), then we obtain the picture
shown in Fig.~\ref{fig:nf3}(d). This shows that the $\epsilon_r(c_I)$
increase monotonically with $k$. This suggests that if we were to vary
$\gamma$, then all three unstable eigenvalues  would appear
first at $k=0$, which is indeed true as can be seen from Fig.~\ref{fig:nf5}(b).
Thus, the marginal state for this profile is again
given by the $k=0$ theory.

\subsection{Finite wavenumber instabilities}
\label{knot0}
The third profile,
\bq
U(y) = y + 5y^3 + f \tanh 4(y-1/2)
,
\label{4.13}
\eq
contains one  or three inflection points, depending on the value
of the parameter $f$. More specifically, as we increase $f$ through
about $1.59$, we create two inflection points near the point
$y=0.6$ (see Fig.~\ref{fig:nf6}). The critical profile for which the two
inflection points emerge contains a point with
$U''=U'''=0$. This leads to a Nyquist plot that contains a
nontransversal intersection of the path with the
$\epsilon_r-$axis; that is, the plot touches the axis but does
not cross it. For larger values of $f$, this degenerate point
splits into the two inflection points and the plot  crosses the
$\epsilon_r-$axis twice. Nyquist plots beyond this bifurcation are shown in
Fig.~\ref{fig:nf6b}(b).

\begin{figure}
\begin{center}
\leavevmode
\epsfysize=7.cm
\epsfbox{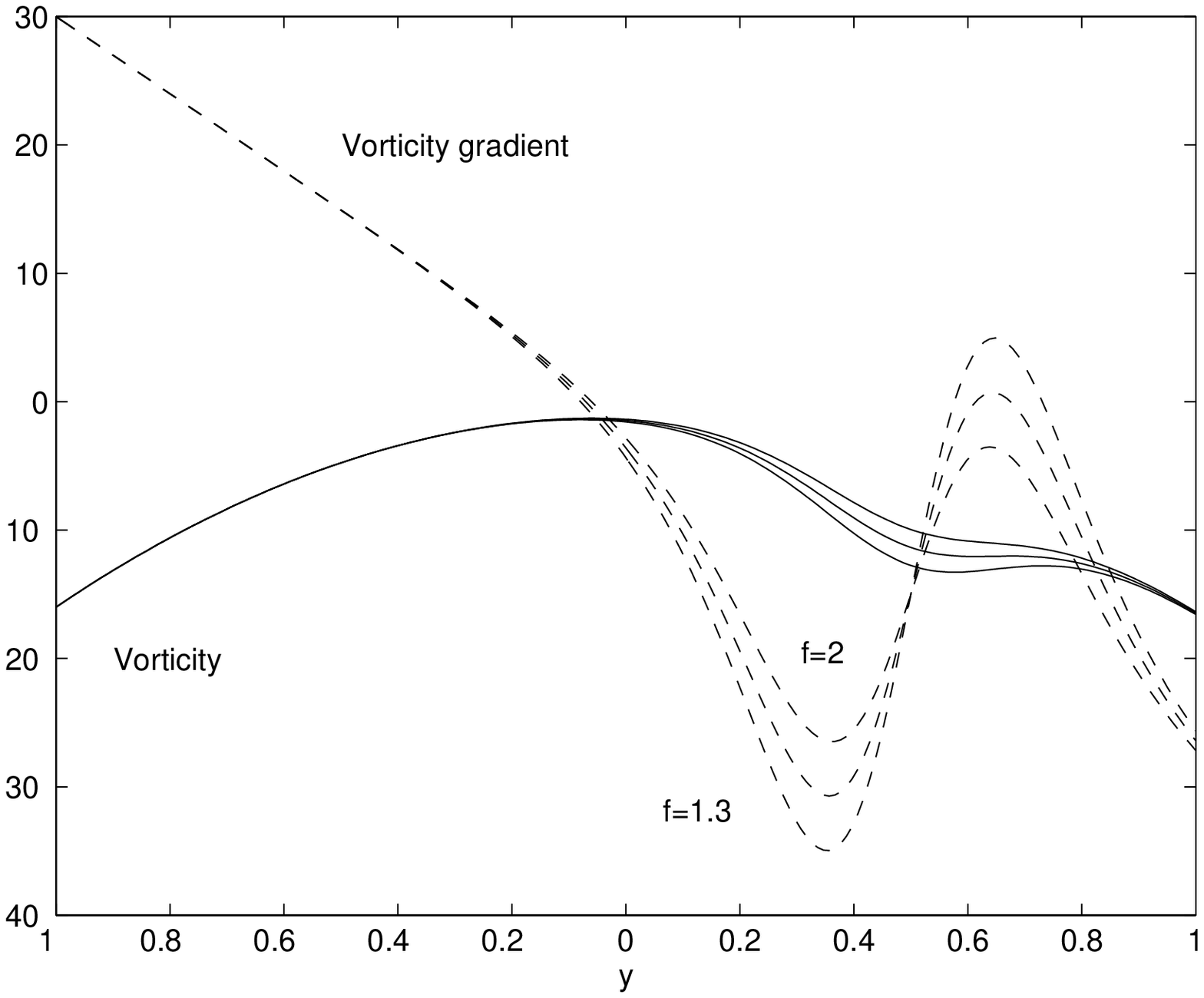}
\end{center}
\caption{Vorticity and vorticity gradient profiles corresponding to the
velocity profile  $U(y) = y + 5y^3 + f \tanh 4(y-1/2)$ with equilibrium parameter values
$f=1.3$, $f= 1.65$, and $f=2$. At $f=f_c\approx1.59$ two inflection points
emerge. }
\label{fig:nf6}
\end{figure}

\iffalse
\begin{figure}
\begin{center}
\leavevmode
\epsfysize=8.cm
\epsfbox{nf44.ps}
\end{center}
\begin{center}
\leavevmode
\epsfysize=6.cm
\epsfbox{nf4b.ps}
\end{center}
\caption{Nyquist plots for $k=0$, $k=0.75$ and $k=1.5$
and $f=1.62$
indicating $k\neq0$ instability onset.}
\label{fig:nf4}
\end{figure}

\fi

\begin{figure}
\begin{center}
\leavevmode
\epsfysize=8.5 cm
\epsfbox{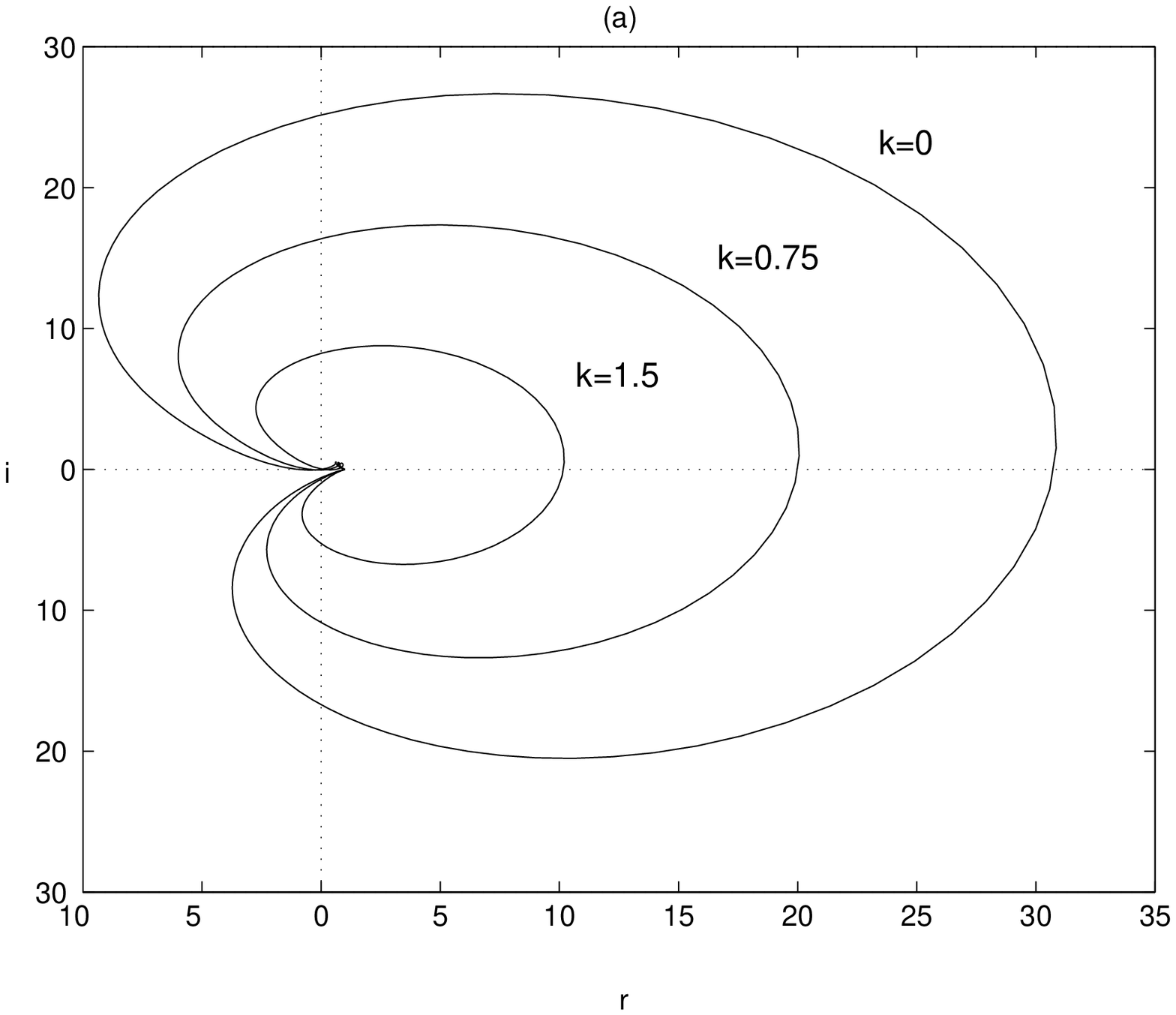}
\end{center}
\end{figure}

\iffalse
\begin{figure}
\begin{center}
\leavevmode
\epsfysize=8.cm
\epsfbox{nf4a.ps}
\end{center}
\end{figure}
\fi

\begin{figure}
\begin{center}
\leavevmode
\epsfysize=8.5 cm
\epsfbox{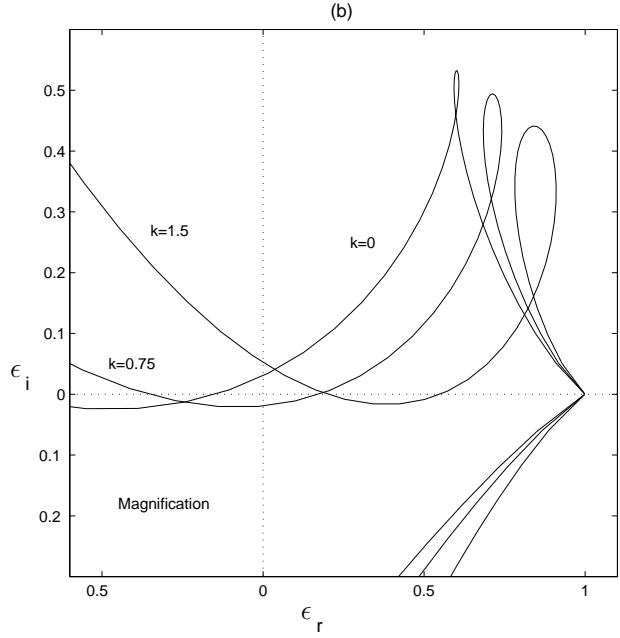}
\end{center}
\caption{(a) Nyquist plots for the  profile  $U(y) = y + 5y^3 + f \tanh
4(y-1/2)$ for $k=0$, $0.75$, and  $1.5$, and $f=1.62$. (b) Magnification that
indicates the onset of instability  through $k\neq0$.}
\label{fig:nf6b}
\end{figure}

\begin{figure}
\begin{center}
\leavevmode
\epsfysize=8.cm
\epsfbox{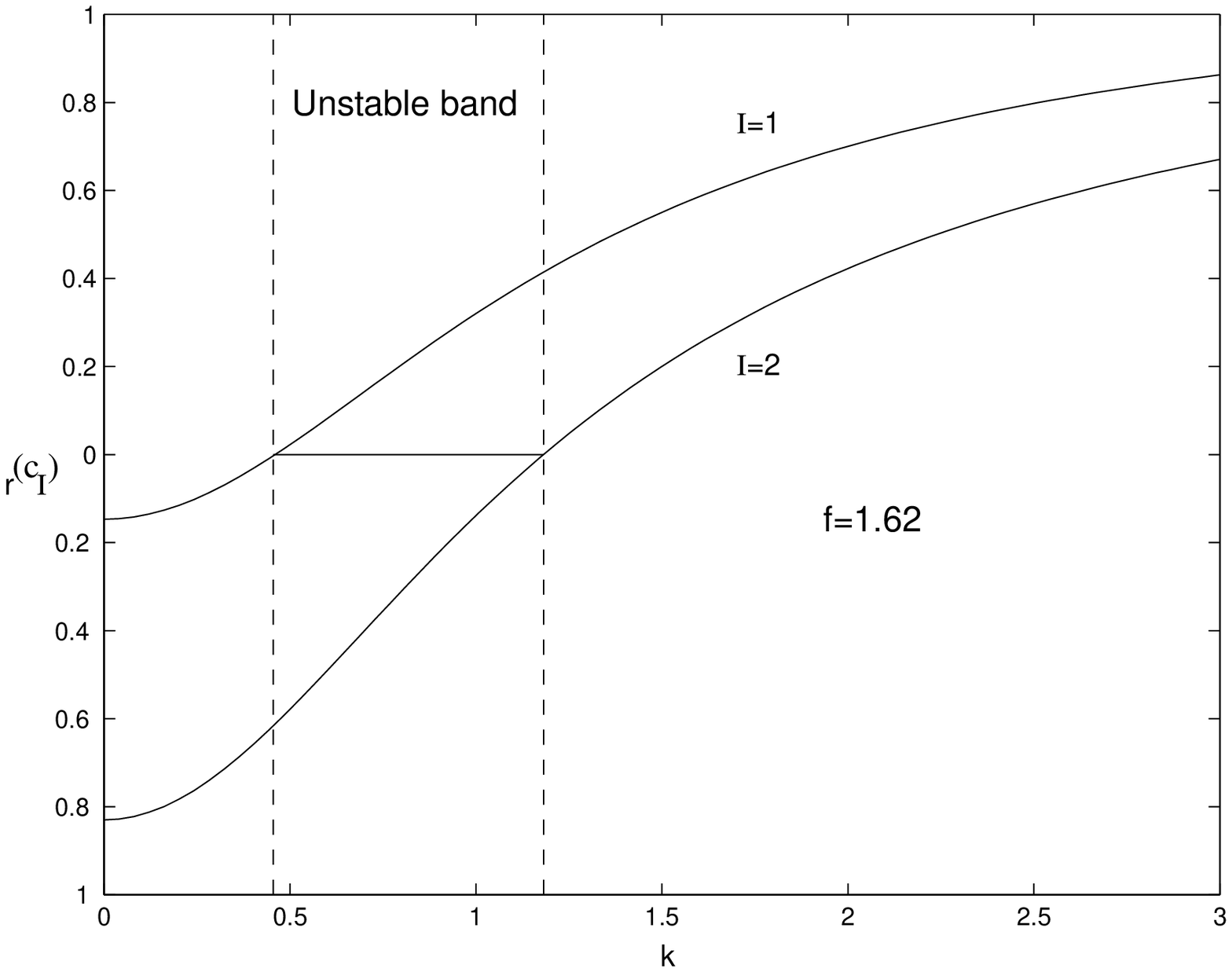}
\end{center}
\caption{Plots of $\epsilon_r(c_I)$ against $k$ at $f=1.62$, where $c_I$,
$I=1,2$, are the two inflection points leading to crossings of
the Nyquist plot in Fig.~6 to the
left of the asymptote $(1,0)$.}
\label{fig:nf4}
\end{figure}

This kind of a change to the profile leads to a situation in
which there can be two crossings of the $\epsilon_r-$axis to
the left of the origin, but in neighborhood of  $k=0$ there are
no unstable eigenvalues. Yet, as we remarked in
Section~\ref{nyq}, increasing $k$ leads to a deformation of the
path  such that the origin is eventually encircled for some
range of $k$ (see Fig.~\ref{fig:nf6b}(a) and (b)). In Fig.~\ref{fig:nf4},
we show
the variation  of $\epsilon_r$ with $k$ for the two newly created inflection
points at $f=1.62$; the range of unstable wavenumbers is $0.12\ltwid k \ltwid
1.31$. Again the variation of $\epsilon_r$ with $k$ is monotonic.

The onset of the unstable band
arises at the value of $f$ for which the two inflection points
emerge; that is, this profile is the marginally stable
state. Moreover, the critical value of $k$ is that required to
make the nontranversal intersection of the path with the
$\epsilon_r-$axis occur at the origin. Thus the emergence of
the two inflection points as we raise $f$ leads to an
instability with an onset at finite wavenumber, as shown
in Fig.~\ref{fig:nf5}(c). Thus, in this case, instability is bounded away from
$k=0$, in contrast to the previous examples. (This particular example
is somewhat analogous to the bump-on-tail instability
in the Vlasov problem \citenum{P60}).

In both Figs.~\ref{fig:nf3}(d)  and \ref{fig:nf4}, $\epsilon_r$ 
appears to be a monotonically increasing function of $k$. If this
feature were generally true, then the deformation of the Nyquist
plots as we vary $k$ would certainly be simpler to
understand. Indeed, this would imply
that $\epsilon_r(c_I)<0$ evaluated at $k=0$ was sufficient
for instability, independently of
the number of inflection points (at least for monotonic,
analytic profiles). However, we have found no argument implying such
a property of $\epsilon_r$, and doubt one exists.

\section{Closing remarks}
\label{con}

We have presented in this paper a  necessary and  sufficient
condition for the instability of monotonic shear flows. Several examples
of equilibrium velocity profiles were treated in Sections \ref{special} and
\ref{num}, demonstrating that the condition  is of practical utility for
finding instability and for understanding the kinds of bifurcations that
can occur.   

One may ask the question of how the method we have presented, which
entails solving a Fredholm integral equation,  compares with directly
calculating the solutions numerically.   Without the Nyquist method, one
could solve Rayleigh's equation in order to locate unstable modes  
with a  given $k$ for $c$ in the complex plane. This amounts to solving a
boundary value problem with singular points. With this procedure one
would   repeatedly solve Rayleigh's  equation at different values of
(complex) $c$ and employ some kind of two-dimensional searching technique
in order to find the eigenvalues. Whilst not especially difficult, this
approach is computationally intensive, and in our opinion is  neither
elegant nor insightful.   As a computational exercise, the calculation of
the Nyquist function is substantially less effort than solving the
boundary-value problem in this fashion. The Nyquist
recipe is  simpler  because  numerically it  only requires  a  matrix
inversion  and only values of  $c$ on the real axis are involved.
Moreover, the sufficient condition  of  ({\ref{3.6})
requires one computation per inflection point, for each value of $k$,
and no search  in the complex plane.

In addition to these computational advantages, the Nyquist method
provides insight into inviscid instability. It allows one to determine the
number of unstable eigenmodes of a profile, as indicated, for example, by
the multiple encircling of the origin of the Nyquist plots of 
Fig.~\ref{fig:nf3}. Also, it leads us to understand the various kinds
of bifurcations to instability of a family of shear flow profiles. Put
another way, Nyquist theory tells us how we can construct profiles with
certain kinds of instabilities. For example, with the Nyquist imagery, we
know how to build profiles that suffer the onset of instability at finite
wavenumber, as in Section 7.3. Without this machinery, it would be much
more difficult even formulating the problem of how to fashion
the needed $U(y)$.

The bifurcations considered here are ones in which unstable eigenvalues 
appear out of a continuous spectrum. It is important to distinguish
this kind of bifurcation from those that occur in systems with only
discrete spectra, since any stable profile is arbitrarily close (in $L_p$
norm) to an unstable one (this is the essence of Gill's result
\citenum{G65} for Couette flow), a feature intimately linked with the
presence of the continuous spectrum. It would then appear that
catagorizing bifurcations to instability is problematic. However, the
Nyquist plots in the
$(\eps_r,\eps_i)-$plane provide a measure of how close a given
equilibrium is from being deformed to one at the onset of instability.
This onset may be through $k=0$ ({\it cf.}\ Figs.~\ref{fig:nf1}
and \ref{fig:nf3})   or at finite wavenumber ({\it cf.}\ 
Fig.~\ref{fig:nf6b}). Indeed, it can occur {\sl via} more degenerate kinds
of bifurcations, and these may be visualized straightforwardly with
Nyquist theory. In fact, it is only the
unclear dependence on $k$ that prevents us from immediately classifying
every instability of a profile with a given number of inflection points
({\it cf.} \citenum{BDY96}).

An example of the kind of insight provided by the Nyquist method is given 
by the following general result: for given $k$, in order to encircle the
origin, the path  must cross the $\epsilon_r-$axis {\sl twice}. That is, in
order to create a new instability we need {\sl two} new inflection points.
Thus, if there are $N$ inflection points in the profile (and so there are
$N+1$ crossing of the $\epsilon_r-$axis altogether), there can be at most
$(N+1)/2$ unstable eigenmodes with that value of $k$. This result, which builds on a theorem
of Howard \citenum{H64}, was stated without proof  in \citenum{F71}.

In finite dimensional Hamiltonian systems bifurcations are regulated
by  Krein's theorem, which states that a necessary condition for
the bifurcation to instability is that colliding eigenvalues possess energy
signatures of opposite sign. It is also know for Hamiltonian systems that
the constancy of energy can be used to obtain  a sufficient  but not
necessary condition for stability. The results of this paper lend insight
to and can be interpreted in  the Hamiltonian context: the sufficient
condition for stability based on energy is equivalent to the conditions of
Rayleigh and  Fj{\o}rtoft,  and a version of Krein's theorem (and
a notion  of signature) for bifurcations involving  the continuous spectra
can be related to the Nyquist function \citenum{BM96}.

We conclude with some remarks about the assumptions we made on the forms
of the profile. The analytic structure of the
Nyquist function heavily relies on the analytic form of the profile in the
vicinity of $[-1,1]$. However, irrespective of the form of $U(y)$, we may
nevertheless construct $\epsilon(c_r)$. In fact, the zeros of this
function for $c_r\in[-1,1]$ still point to the existence of
regular neutral eigenmodes even when $U(y)$ is not analytic.
What is no longer clear is how to continue $\epsilon$ off the real axis.
Or equivalently, whether there are discrete complex modes at
nearby parameter values. Nevertheless there are various hints
in this work that suggest that the theory works even if $U(y)$
has only two derivatives.

Finally, nonmonotonicity of $U(y)$ brings in the
new complications of multiple critical layers for a
given wave speed, and unusual properties of
the streamfunction for critical layers lying at the
shearless points where $U'(y)=0$. It is for
these reasons that the theory does not immediately carry
over. We are currently working on these and other extensions.

\section*{Acknowledgements}
This work was supported by the U.S. DoE under contract
No.~DE-FG05-80ET-53088. We thank J. B. Keller for
commenting on the manuscript, and S. Friedlander for drawing our attention
to reference \citenum{F71}.

\section*{Appendix}
\label{Appendix A}

In this appendix we show that the change in the argument of the Wronskian
is determined entirely by the piece of the contour $\cac$ that is along
the flow domain ($5\rightarrow 7$ of Fig.~\ref{fig:nf2}); the remainder of the
contour will be shown to not encircle the origin.

\subsection*{The big semicircle: $1\rightarrow 2\rightarrow 3$.}

On the big semicircle $c=Re^{i\theta}$ with $\theta\in(0,\pi)$. We
consider the limit  $R=|c|\rightarrow\infty$ by invoking the principle of
permanence (see {\it e.g.}~\citenum{H76}), which in the present context
simply states that the limit $c\rightarrow\infty$ of the solution to
Rayleigh's equation at fixed $y$ is equal to the solution of the
$c\rightarrow\infty$ limit of Rayleigh's equation. The latter limit is
$\psi'' -k^2\psi=0$, which has the following solution with the appropriate
boundary conditions: $\Phi^{\infty}_<(y):= \sinh[k(y+1)]/k$.
The  principle of permanence implies
$\Phi_<(y,c\rightarrow\infty)=\Phi^{\infty}_<(y)$,
and thus using $W(c)= -\Phi_<(1,c)$ we obtain
$\lim_{R\rightarrow\infty}W(c)=-\sinh(2k)/k<0$.
Therefore, the piece of the contour  $1\rightarrow
2\rightarrow 3$  maps into the $W$-plane as a single point on the negative
real axis (this is the asymptote $(1,0)$ on the $\epsilon-$plane).

As a check consider the limit  $k\rightarrow 0$. Observe $\lim_{k\rightarrow0}
\Phi^{\infty}_<(1,c)=2$. (This is clearly correct  since  Rayleigh's
equation  becomes $\psi''=0$   and the assumed boundary conditions imply
$\psi=1+y$, which when evaluated at $y=1$ gives 2\@.) Defining
$F(c):= W(c)/[(U_2-c)(c-U_1)]$ gives $F(c,0)\sim  2e^{-2i\theta}/ R^2$,
which is consistent with  the $k=0$ result of \citenum{RS64}.

The principle of permanence can be demonstrated explicitly by using
the solution to Rayleigh's equation   written as
$\psi(y,c)=\sum_{n=0}^\infty\psi_n(y,c)$,
where
\begin{equation}
\psi_{n+1}(y,c)=\int^y_{-1}(y-y')\left(k^2
+\frac{U''(y')}{U(y')-c}\right)
\psi_n(y',c)\,dy'\,.
\label{int.it}
\end{equation}
Using Cauchy's inequality, it is not difficult to  prove that this series
converges uniformly for all $y\neq y_c$. If we choose $\psi_0$ so that
the boundary conditions of (2) are satisfied, evaluate (\ref{int.it}) at
$y=1$, and slip the $c\rightarrow \infty$ limit through the integral sign,
then
we obtain $\Phi^{\infty}_<(y)$.

As an aside, note that each $\psi_n$ is
analytic in $c$.  Thus because of uniform convergence,   $\psi(1,c)$ is
analytic  for  $c\neq U(1)$.   This is true even for  profiles $U(y)$ that are
not analytically continuable into the complex plane; it is only necessary for
the integral in (\ref{int.it}) to exist to get analyticity in $c$. This puts a
relatively  mild restriction on $U$. For example, if $c$ is not in the flow
domain, then $y U''(y)\in L_1[-1,1]$ is sufficient.

\subsection*{The exterior legs: $3\rightarrow 4$ and $8\rightarrow 1$.}

We show below that if $c$ is not in the flow domain, which is the case on the
legs $3\rightarrow 4$ and $8\rightarrow 1$   neither  $\Phi_<(y,c)$ nor
$\Phi_>(y,c)$ can vanish.   Evaluating  $\Phi_<$ at $y=1$ we see
the same is true for $W$.  We also include a proof that $W$ is a monotonic
functions of $c$  on $3\rightarrow 4$ and $8\rightarrow 1$. Thus these
pieces of
the contour map into curves that   cannot cross into the right hand portion of
the $W$-plane.

To prove  the above statements  we use a formula introduced  by
Green  in the first half of the nineteenth  century (see  {\it e.g.}\
\citenum{H76}  or \citenum{I44}). For Rayleigh's equation, Green's formula
is
\bq
\left[
\psi\,\left(\psi' -  \frac{U'\, \psi}{U-c}\right)
\right]_{y_0}^y
= \int_{y_0}^{y} \left[ \left(\psi' -  \frac{U'\, \psi}{U-c}\right)^2
+ k^2 \, \psi ^2
\right] dy'>0 \,,
\label{Green}
\eq
which is valid for any solution $\psi$. This formula can be derived from
Rayleigh's equation by  multiplying by $\psi$, manipulating, and integrating.
It is important to remember that $c$ is assumed to be real and outside the
domain of integration.

Upon taking $y_0=-1$, $\psi=\Phi_<$, and  applying the boundary condition
$\Phi_<(-1,c) =0$, (\ref{Green}) implies
\bq
\Phi_<(y,c)\,\left[\Phi_<'(y,c) -  \frac{U'(y)\,
\Phi_<(y,c)}{U(y)-c}\right] >0\,,
\label{green.inequal}
\eq
for all $y\in (-1,1]$  and $c\notin[U_1,U_2]$. This inequality means that
neither factor can vanish for $y$ in the interior of the flow domain.
We know that $\Phi_<(-1,c)=0$ and that  $\Phi'_<(-1,c)=1$, and  therefore in a
neighborhood of $y=-1$, by continuity of the solution, $\Phi_<(y,c)>0$.
Thus $\Phi_<(y,c)>0$ for all $y\in (-1,1]$, and
inequality (\ref{green.inequal})  implies that the $[~~]$-factor must
also be positive. (Note that neither factor can  be singular by the
existence theorem applied to  Rayleigh's  equation\@.) We mention, for
later use, that a similar   argument   shows    $\Phi_>(y,c)>0$.

Evaluating $\Phi_<$ at $y=1$ yields $W(c)=-\Phi_<(1,c)<0$ for all
$c\in(-\infty,U_1)$, which is our desired result for the leg
$3\rightarrow 4$.  Similarly,   $W(c)<0$ for all $c\in(U_2,
\infty)$,   our desired result for the leg $8\rightarrow 1$.

We now further demonstrate that $W$ is monotonic, although this
is not strictly needed for the proof.  To this end we differentiate Rayleigh's
equation with respect to $c$, yielding
\bq
 \frac{\p \psi''}{\p c} -
\left(k^2 + \frac{U''(y)}{U(y)-c}\right)\frac{\p \psi}{\p c}
=\frac{U''(y)}{[U(y)-c]^2}\,
\psi\,,
\label{Ray.c}
\eq
which   by the method of variation of parameters, is seen to have the
following solution:
\bq
\frac{\p \psi(y,c)}{\p c}= \int_{-1}^{y}\frac{\psi(y',c)}{W_{12}}
\frac{U''(y')}{[U(y')-c]^2}\,
\Big[\psi_1(y',c)\psi_2(y,c)-\psi_2(y',c) \psi_1(y,c)\Big]\, dy'\,.
\label{Ray.c.int}
\eq
Here $\psi_1$ and $\psi_1$ are any two independent solutions of Rayleigh's
equation and
\bq
W_{12}(c):=\Big[\psi_1(y,c)\psi_2'(y,c)-\psi_2(y,c)
\psi_1'(y,c)\Big]\,.
\label{W12}
\eq
Observe that (\ref{Ray.c.int})   satisfies $\p \psi(-1,c)/\p
c=0$,  which is consistent with $\psi(-1,c)=0$, and thus we may assume
$\psi(y,c)=\Phi_<(y,c)$. Letting $\psi_1(y,c)= \Phi_<(y,c)$ and
$\psi_2(y,c)=\Phi_>(y,c)$ gives
\bq
W_{12}(c):=[\Phi_<(y,c)\Phi_>'(y,c)-\Phi_>(y,c)
\Phi_<'(y,c)\Big]= -\Phi_>(-1,c) = \Phi_<(1,c)\,,
\label{15}
\eq
and the expression  (\ref{Ray.c.int}) implies
\bq
\frac{\p  \Phi_<(1,c)}{\p c}=-\int_{-1}^{1}
\frac{U''(y')}{[U(y')-c]^2}\,
 \Phi_<(y',c)\,\Phi_>(y',c) \,
dy'\,.
\label{Phi.c}
\eq
Integrating (\ref{Phi.c}) by parts gives
\bqy
\frac{\p  \Phi_<(1,c)}{\p c} &=&\int_{-1}^{1} \left\{
 \frac{-2\,U'^2}{[U(y')-c]^3}\,\Phi_<(y')\,\Phi_>(y') \right.
\nonumber\\
&+& \left. \frac{U'}{[U(y')-c]^2}\,\left[ \Phi'_<(y')\,\Phi_>(y')
     + \Phi_<(y')\,\Phi'_>(y')\right]\right\}\,
dy'\,,
\label{Phi.c.1}
\eqy
which upon insertion of  (\ref{15}) into its second term
can be manipulated into
\bq
\frac{\p  \Phi_<(1,c)}{\p c} = K(c) + \mu(c)\, \Phi_<(1,c)\,,
\eq
where
\bq
\mu(c):= \int_{-1}^{1}\frac{U'(y')}{[U(y')-c]^2}\, dy'
 = \frac{U_2-U_1}{(U_1-c)(U_2-c)}\,.
\label{mu}
\eq
and
\bq
K(c):=\int_{-1}^{1} \left\{
 \frac{2\,U'(y')\,\Phi_>(y')}{[U(y')-c]^4}\,
\left[
\Phi'_<(y')  - \frac{U'(y') \,\Phi_<(y')}{U(y')-c}
\right]
\right\}\,
dy'    \,.
\label{K}
\eq
Clearly for $c\notin(U_1,U_2)$, $\mu(c)>0$.  The $[~~]$-factor of (\ref{K})
is precisely   the $[~~]$-factor of (\ref{green.inequal}), which we showed
is positive. Since   $U'(y)>0$  and $\Phi_>(y)>0$, for $y\in[-1,1)$, we see
that $K(c)>0$.  Thus we have established monotonicity:
$\p \Phi_<(1,c,k )/\p c >0$  or $\p W(c,k )/\p c <0$ for
$c\notin(U_1,U_2)$.

\subsection*{The boundary regions: $4\rightarrow 5$ and $7\rightarrow 8$.}

Now we consider the pieces of the contour that skirt the boundaries of the
flow domain. These are the pieces  $4\rightarrow 5$ and
$7\rightarrow 8$ of Fig.~\ref{fig:nf2}.  We use the expressions $\psi_g$ and
$\psi_b$ of (\ref{2.6}) and (\ref{2.7}) to construct $\Phi_<(y,c)$ according to
the procedure described in  Section~\ref{disp.rel}. Upon enforcing the boundary
conditions as described in  that section, it is not difficult to show that the
Wronskian has the following form:
\bqy
W(c) &=& - \Phi_<(1,c) = \chi_1(c) (U_1-c)+ \chi_2(c) (U_2-c)
\nonumber\\
 &+ & (U_2-c) (U_1-c)[\chi_3(c)\, \ln(U_1-c)
 + \chi_4(c)\, \ln(U_2-c)]\,,
\label{W.chi}
\eqy
where the functions $\chi_i$ for $i=1-4$ are analytic functions of $c$.
The piece $4\rightarrow 5$ can be pulled down into a little semicircle on
which
$c=U_1 + \delta e^{i\theta}$,  where $\delta >0$ and $\theta\in(-\pi,0)$,
which
is still consistent with $c$ being in the upper half plane.  Evidently,
$\lim_{\delta\rightarrow 0}W(c)  =  \chi_2(U_1)(U_2-U_1)$.  Since this piece of
the contour must connect to the piece that terminates at 4 as $3\rightarrow 4$,
it follows that the piece of the contour  $4 \rightarrow 5$  maps
into a point on the negative  real axis of the  $W$-plane. Similarly, on
$7\rightarrow 8$, where   $c=U_2 + \delta e^{i\theta}$,  $\delta >0$,  and
$\theta\in(-\pi,0)$, we obtain
$\lim_{\delta\rightarrow 0}W(c)  = \chi_1(U_2)(U_1-U_2)$,
and   the piece of the contour  $7 \rightarrow 8$  maps into a point on the
negative real axis of the  $W$-plane.

So in conclusion, since  we have proven that for $c\in\calc'$ the real part of
$W$ is negative, the only part  of the contour that can give rise to a
change in the argument of $W$ is the piece from $5\rightarrow 7$.

\bigskip

\centerline{UNIVERSITY OF NOTTINGHAM}
\centerline{UNIVERSITY OF TEXAS AT AUSTIN}

\clearpage

\section*{Figure Captions:}

\begin{description}

\item[Figure 1:]
Nyquist plots for the single inflection point profile,
$U(y)=\tanh\beta y$. (a) Four plots for $\beta=2$ and $k=0$, $1$, and
$2$,   and  the critical value for the onset of instability,
$k=k_c \simeq 1.832$ (dashed curve). (b) Four plots for $k=0$ and
$\beta=1$, $1.5$, and $2$,  and  the critical value for the onset of
instability,
$\beta=\beta_c \simeq 1.2$ (dashed curve).

\item[Figure 2:]
The contours $\calc$, $\calc'$, and $H$ in the
$c$-plane. The closed contour $\calc$  runs the entire circuit  from
$1\rightarrow 2\rightarrow\dots \rightarrow 8\rightarrow 1$, with the
portion from $3\rightarrow1$ lifted  infinitesimally above the real
axis. The contour  $\calc'$ is   $\calc$ with the piece along the flow
domain, $5\rightarrow7$,  removed. The contour $H$ (dashed) is Howard's
semicircle, within which the unstable eigenvalues must lie.

\item[Figure 3:]
Stability boundaries for the three equilibrium profiles of Sec.~7.
Figures (a) and (b) depict the nucleation of instability through $k=0$,
while (c) depicts the nucleation of instability through $k\neq 0$.
In (b), the stability boundary of the second and third unstable modes
is also shown.

\item[Figure 4:]
Nyquist plots for the   multiple inflection point  profile,
$U(y)=y + (\sin\gamma y)/2\gamma$, with $k=1$. (a) The case $\gamma=5$  has
one encircling of the origin and thus one unstable eigenvalue. (b) In the case
with $\gamma=7.5$ two more loops have appeared and the
curve almost encircles the origin two more times. That is, it is
close to the threshold of the instability of two further modes.
(c) The case $\gamma=10$ has three encirclings
of the origin and three unstable eigenvalues. (d) Depiction of the monotonic
dependence of the crossing values to the left of the asymptote (1,0)  as a
function of $k$.

\item[Figure 5:]
Vorticity and vorticity gradient profiles corresponding to the velocity
profile  $U(y) = y + 5y^3 + f \tanh 4(y-1/2)$ with equilibrium parameter
values $f=1.3$, $f=1.65$, and $f=2$. At $f=f_c\approx1.59$ two inflection
points emerge.

\item[Figure 6:]
(a) Nyquist plots for the  profile $U(y) = y + 5y^3 + f \tanh
4(y-1/2)$ for $k=0$, $0.75$, and  $1.5$, and $f=1.62$. (b) Magnification that
indicates the onset of instability  through $k\neq0$.

\item[Figure 7:]
Plots of $\epsilon_r(c_I)$ against $k$ at $f=1.62$, where $c_I$,
$I=1,2$, are the two inflection points leading to crossings of
the Nyquist plot in Fig.~6 to the
left of the asymptote $(1,0)$.

\end{description}

\end{document}